\documentclass[acmsmall]{acmart}
\usepackage{url}
\usepackage{graphicx} % Required for inserting images
\usepackage[utf8]{inputenc}
\usepackage{amsmath}
\usepackage{amsfonts}
\usepackage{aliascnt}
\usepackage{amsthm}
\usepackage{bussproofs}
\usepackage{dirtytalk}
\usepackage{float}
\usepackage{xspace}
\usepackage{hyperref}
\usepackage{xifthen}
\usepackage[nomap]{FiraMono}
\usepackage{tikz}
\usepackage{stmaryrd}
\usepackage{subcaption}
\usepackage{framed}
\usetikzlibrary{positioning}
\usepackage[ruled,vlined,linesnumbered]{algorithm2e}
    \SetAlFnt{\fontsize{8.5}{10}\selectfont}
    \SetArgSty{textup}
    \SetProgSty{textup}
    \SetFuncArgSty{textup}
    \SetFuncSty{textit}
    \SetKwProg{procedure}{def}{}{}
    \SetKwProg{datastructure}{Datastructure}{}{}
    \SetKwSwitch{Switch}{Case}{Other}{match}{:}{case}{otherwise}{endcase}{endsw}
    
    \SetKw{KwCont}{continue}
    \SetKw{KwBreak}{break}
    
    \newcommand{\True}{\texttt{True}}
    \newcommand{\False}{\texttt{False}}
    
\newcommand{\lineIf}[2]{\textbf{if}~{#1}~\textbf{then}~{#2}\;}
\newcommand{\lineElseIf}[2]{\textbf{else if}~{#1}~\textbf{then}~{#2}\;}

\SetKwComment{Comment}{/* }{ */}

\newcommand{\A}{\mathcal A}
\newcommand{\B}{\mathcal B}

\theoremstyle{definition}

\newtheorem{thm}{Theorem}[section]

\newaliascnt{lemma}{thm}
\newtheorem{lemma}[lemma]{Lemma}
\aliascntresetthe{lemma}

\newaliascnt{cor}{thm}
\newtheorem{cor}[cor]{Corollary}
\aliascntresetthe{cor}

\newaliascnt{defin}{thm}
\newtheorem{defin}[defin]{Definition}
\aliascntresetthe{defin}

\definecolor{green}{rgb}{0, 0.6, 0}

\definecolor{comments}{RGB}{80,0,110}

\newcommand{\OL}{{O{\kern-0.1em}L}}
\newcommand{\POL}{{P{\kern-0.1em}O{\kern-0.1em}L}}
\newcommand{\FOL}{{F{\kern-0.1em}O{\kern-0.1em}L}}
\newcommand{\CL}{{C{\kern-0.1em}L}}

\newcommand{\FOLm}{\mathit{F{\kern-0.1em}({\kern-0.1em}O{\kern-0.1em}L{\kern-0.1em})\textsuperscript{2}}}

\newcommand{\ssimw}[1]{{{\kern-0.05em}/{\kern-0.08em}{#1}}}

\newcommand{\sime}{\sim_{{\kern-0.1em}E}}
\newcommand{\siml}{\sim_{{\kern-0.1em}L}}

\newcommand{\eqclass}[2]{{[#1]}_{#2}}
\newcommand{\eqclasssim}[2]{\eqclass{#1}{\sim_{{\kern-0.1em}#2}}}

\newcommand{\term}[2]{\mathcal{T}_{#2}%
    \ifthenelse{\isempty{#1}}%
    {}%
    {(#1)}%
}
\newcommand{\hypname}{Hyp}
\newcommand{\relevantS}[1]{\textsf{relevant}({#1})}
\newcommand{\sizeOf}[1]{\|{#1}\|}

\newcommand{\EPROLD}{\textsf{EPR-OL-D}}

\setcopyright{rightsretained}
\acmDOI{10.1145/3632881}
\acmYear{2024}
\copyrightyear{2024}
\acmSubmissionID{popl24main-p183-p}
\acmJournal{PACMPL}
\acmVolume{8}
\acmNumber{POPL}
\acmArticle{39}
\acmMonth{1}
\received{2023-07-11}
\received[accepted]{2023-11-07}


\begin{document}

\title{Orthologic with Axioms}
\author{Simon Guilloud}
\authornote{We acknowledge the financial support of the Swiss National Science Foundation project 200021\_197288
``A Foundational Verifier''. The paper is available from ACM, from the authors, as well as from \url{https://doi.org/10.48550/arXiv.2307.07569}}
\orcid{0000-0001-8179-7549}
\affiliation{%
  \institution{EPFL, IC}
  \city{CH-1015 Lausanne}
  \country{Switzerland}
}
\email{simon.guilloud@epfl.ch}

\author{Viktor Kun\v{c}ak}
\authornotemark[1]
\orcid{0000-0001-7044-9522}
\affiliation{%
  \institution{EPFL, IC}
  \city{CH-1015 Lausanne}
  \country{Switzerland}
}
\email{viktor.kuncak@epfl.ch}

\begin{abstract}
      We study the proof theory and algorithms for orthologic, a logical system based on ortholattices, which have shown practical relevance in simplification and normalization of verification conditions. Ortholattices weaken Boolean algebras while having polynomial-time equivalence checking that is sound with respect to Boolean algebra semantics. We generalize ortholattice reasoning and obtain an algorithm for proving a larger class of classically valid formulas.

    As the key result, we analyze a proof system for orthologic augmented with axioms. An important feature of the system is that it limits the number of formulas in a sequent to at most two, which makes the extension with axioms non-trivial. We show a generalized form of cut elimination for this system, which implies a sub-formula property. From there we derive a cubic-time algorithm for provability from axioms, or equivalently, for validity in finitely presented ortholattices. We further show that propositional resolution of width 5 proves all formulas provable in orthologic with axioms. We show that orthologic system subsumes resolution of width 2 and arbitrarily wide unit resolution and is complete for reasoning about generalizations of propositional Horn clauses. 
    
    Moving beyond ground axioms, we introduce effectively propositional orthologic (by analogy with EPR for classical logic), presenting its semantics as well as a sound and complete proof system. Our proof system implies the decidability of effectively propositional orthologic, as well as its fixed-parameter tractability for a bounded maximal number of variables in each axiom. As a special case, we obtain a generalization of Datalog with negation and disjunction.

\end{abstract}
\maketitle
\sloppy

\section{Introduction}
Our goal is to construct efficient building blocks for theorem proving and program verification. coNP-hardness of propositional logic already presents a barrier to large-scale reasoning, such as simplification of large formulas and using intermediate assertions to help software verification. We aim to improve the worst-case efficiency of reasoning while preserving the spirit of a specification language with conjunction, disjunction, and negation. We therefore investigate the concepts of \emph{ortholattices} and \emph{orthologics} as a basis of predictable reasoning.

Non-distributive generalizations of classical logic, including orthologic, were  introduced as \textit{quantum logic} to describe experiments in quantum mechanics, where it was realized that distributivity fails \cite{birkhoffLogicQuantumMechanics1936, bellOrthologicForcingManifestation1983}. The term \textit{orthologic} was used in \citep{goldblattSemanticAnalysisOrthologic1974} for the logic corresponding to the algebraic class of ortholattices, a generalization of Boolean algebras. 
In particular, the class of closed subsets of a Hilbert space is an ortholattice, but not a Boolean algebra \cite{rawlingOrthologicQuantumLogic2000}. In theoretical physics, ortholattices are an intermediate step towards the study of \textit{orthomodular lattices} and \textit{modular ortholattices} among others, \citep{hardegreeMaterialImplicationOrthomodular1981, kalmbachOrthomodularLattices1983, sherifDecisionProblemOrthomodular1997, hyckoImplicationsEquivalencesOrthomodular2005}. 
Ortholattices have also found application in modelling of epistemic modal logic \citep{hollidayOrthologicEpistemicModals2022, hollidayFundamentalNonClassicalLogic2023}.
Recently, researchers have proposed to use ortholattices as an efficient approximation to classical logic in automated reasoning, in the context of both proof assistants and software verifiers.

In the domain of proof assistants, researchers have recently used orthologic in the design of a proof assistant named LISA \citep{guilloudLISAModernProof2023, GuilloudETALLisa}. LISA's proof checker supports classical sequent calculus, but checks the assumptions and the conclusion of each step modulo orthologic equivalence. This approach retains polynomial-time checking for proofs, which is important for scalability. At the same time, it allows a single step to, for example, reorder arguments of conjunctions and disjunction, use idempotence, absorption and negation normal form transformations.

In the domain of program verifiers, a recent approach \cite{guilloudFormulaNormalizationsVerification2023} uses a normal-form algorithm for ortholattices to improve the cache hit ratio of verification conditions. It also serves to simplify and shorten formulas before sending them to a general but potentially costly SMT solver.

The results suggest that ortholattices can be used to simplify and check equivalence of large formulas using polynomial-time algorithms, while providing both soundness and a clear mental model of the degree of its incompleteness relative to Boolean algebra reasoning.
We expect that there are other applications in the domain of programming languages, and possibly more broadly in computer science, for which long runtime and non-determinism leading to difficulties in reproducing results are major drawbacks, and hence where predictability and efficiency at the expense of classical completeness can be a desirable trade-off. 

In the domain of type systems, researchers have recently revisited Boolean algebra as a component of principled approaches for expressive type systems \cite{parreauxMLstructPrincipalType2022}. Whereas most questions in Boolean algebras have no known deterministic polynomial-time algorithms, using orthologic instead of Boolean algebra could lead to faster, polynomial-time, reasoning about such situations. Being able to use axioms in orthologic inference, which is our key contribution, permits including knowledge of pre-existing subtyping and inhabitation relations in the typing context.
Negated types are also relevant in languages with pattern matching: if a term $t$ fails to match a type $T$, it is possible to deduce that $t: \neg T$. Such precedence-based reasoning further appears in the match types of Scala \cite{blanvillainTypelevelProgrammingMatch2022, blanvillainAbstractionsTypeLevelProgramming2022}.
Type checking procedures for programming language with variants of liquid and refinement types \citep{vazouRefinementTypesHaskell2014, freemanRefinementTypesML1991} may also benefit from orthologic reasoning. 

Extensions of Datalog with a negation operator \cite{clarkNegationFailure1978, kunenNegationLogicProgramming1987} is also a suitable candidate for employing orthologic reasoning, potentially leading to more expressive program analyses encoded using Datalog rules \cite{DBLP:conf/pldi/WhaleyL04}. We set the foundation for such directions with our Datalog-related results in \autoref{sec:epr}.

To motivate the new results of this paper, note that an ortholattice algorithm in \cite{guilloudFormulaNormalizationsVerification2023} may reduce $x \land z \land \lnot (u \land \lnot x)$ to the normal form $x \land z$.
This normal form is based on the laws that hold in \emph{all} ortholattices (equivalently, in the free ortholattice). This makes the technique widely applicable, but it also makes it weak in terms of classical and domain-specific tautologies it can prove. 
%It has been shown how the $\leq$ relation on ortholattices can be used to encode properties of underlying theories, for example of arrays, integers or equality, but without model-theoretic completeness characterization.
The present paper explores making orthologic-based reasoning more precise and more usable, asking the following questions:
\begin{enumerate}
    \item Can we formally extend orthologic with ground non-logical axioms between propositions?
    \item Can we find a complete and efficient algorithm for solving such problems?
    \item What are classes of formulas in classical logic for which orthologic proofs always exist?
    \item Can orthologic (with axioms) be used beyond propositional logic, e.g., for predicate logic?
\end{enumerate}
Our approach to these questions is to use a sound and complete proof system for orthologic that we extend to support arbitrary non-logical axioms. Algebraically, our proof system is complete for establishing inequalities in the class of ortholattices specified by a given \emph{presentation}, a set of ortholattice inequalities.
From the practical point of view, using axioms to represent part of the input formula gives a sound and strictly stronger approximation of classical logic than using ortholattices without axioms. 

\subsection{Ortholattices}

\begin{figure}[ht]
    \centering

    \fbox{
    \begin{tikzpicture}[
        roundnode/.style={circle, very thick, minimum size=5mm},
        ]
        %Nodes
        \node[roundnode]      (l)                              {$1$};
        \node[roundnode]        (b)       [below left=0.6cm of l] {$b$};
        \node[roundnode]      (a)       [below=0.6cm of b] {$a$};
        \node[roundnode]        (na)       [below right=0.6cm of l] {$\neg a$};
        \node[roundnode]      (nb)       [below=0.6cm of na] {$\neg b$};
        \node[roundnode]        (o)       [below right=0.6cm of a] {$0$};
        
        %Lines
        \draw[-] (l) -- (b);
        \draw[-] (b) -- (a);
        \draw[-] (a) -- (o);
        \draw[-] (l) -- (na);
        \draw[-] (na) -- (nb);
        \draw[-] (nb) -- (o);
        \end{tikzpicture}}
\qquad
        \fbox{\begin{tikzpicture}[
        roundnode/.style={circle, very thick, minimum size=5mm},
        ]
        %Nodes
        
        \node[roundnode]      (main)                              {};
        \node[roundnode]      (l)         [above=0.5cm of main] {$1$};
        \node[roundnode]      (o)         [below=0.5cm of main] {$0$};
        \node[roundnode]        (na)       [left=0.2cm of main] {$\neg a$};
        \node[roundnode]        (a)       [left=0.5cm of na] {$a$};
        \node[roundnode]        (b)       [right=0.2cm of main] {$b$};
        \node[roundnode]        (nb)       [right=0.5cm of b] {$\neg b$};

        %Lines
        \draw[-] (l) -- (a);
        \draw[-] (l) -- (na);
        \draw[-] (l) -- (b);
        \draw[-] (l) -- (nb);
        \draw[-] (a) -- (o);
        \draw[-] (na) -- (o);
        \draw[-] (b) -- (o);
        \draw[-] (nb) -- (o);
        \end{tikzpicture}}
    \caption{Ortholattices $O_6$ and $M_4$. An ortholattice is distributive iff if it contains neither as a sub-ortholattice.}
    \label{fig:O6M4}
\end{figure}
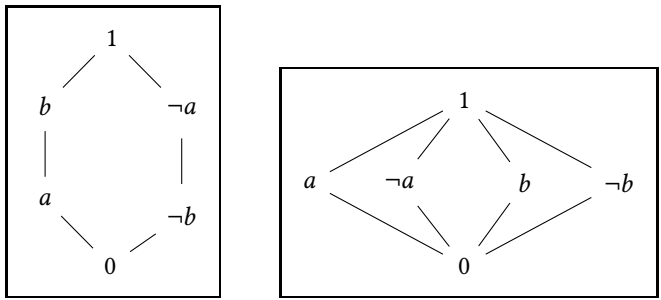

Ortholattices are a weaker structure than Boolean algebras, where distributivity does not necessarily hold. \autoref{tab:algebraiclaws} shows their axiomatization. All Boolean algebras are ortholattices; they are precisely those ortholattices that are distributive. \autoref{fig:O6M4} shows two characteristic finite non-distributive ortholattices; keeping these structures in mind may provide intuition for reasoning inside the class of all ortholattices.
\emph{Orthologic} is the logical system that corresponds to ortholattices, analogously to how classical logic corresponds to Boolean algebras (and intuitionistic logic to Heyting algebras).
\begin{table}[bth]
    \centering
    \begin{tabular}{r c @{\hskip 2em} | @{\hskip 2em} r c}
         V1: & $x \lor y = y \lor x$  & V1': & $x \land y = y \land x$ \\
         V2: & $x \lor ( y \lor z) = (x \lor y) \lor z$  & V2': & $x \land ( y \land z) = (x \land y) \land z$ \\
         V3: & $x \lor x = x$  & V3': & $x \land x = x$ \\
         V4: & $x \lor 1 = 1$  & V4': & $x \land 0 = 0$ \\
         V5: & $x \lor 0 = x$  & V5': & $x \land 1 = x$ \\
         V6: & $\neg \neg x = x$\\
         V7: & $x \lor \neg x = 1$  & V7': & $x \land \neg x = 0$ \\
         V8: & $\neg (x \lor y) = \neg x \land \neg y$  & V8': &  $\neg (x \land y) = \neg x \lor \neg y$ \\
         V9: & $x \lor (x \land y) = x$ & V9': & $x \land (x \lor y) = x$   \\
    \end{tabular}
    
    \
    
    \caption{Laws of ortholattices, algebraic varieties with signature $(S, \land, \lor, 0, 1, \neg)$ from \cite{guilloudFormulaNormalizationsVerification2023}. These non-minimal laws illustrate the duality between $\land$ and $\lor$ and have explicit bottom and top element. 
    See, e.g., \cite{DBLP:journals/ipl/McCune98} for an equivalent smaller axiomatization.
    \label{tab:algebraiclaws}}
\end{table}

\subsection{Example of Using Axioms}
\label{ex:intro}

Using axioms in orthologic inference allows us to prove more classical implications than by encoding the entire problem into one formula, increasing the power of reasoning. To understand why, note that proving validity of an implication $L \rightarrow R$ in all ortholattices can be phrased as proving $L \le R$ in all ortholattices, for all values to which $L$ and $R$ can evaluate in those ortholattices. Such an inequality needs to hold in the ortholattice $O_6$ in \autoref{fig:O6M4} when $L$ evaluates to, for example, $b$. On the other hand, using axioms, we can encode an implication problem as follows: prove that, in every ortholattice, if $L=1$, then also $R=1$. Because $L$ is restricted to be $1$, what remains to prove is a weaker statement, provable for more formulas. The conclusion remains sound with respect to the $\{0,1\}$ lattice of classical logic, where $L=1$ is the only non-trivial case to check for inequality. 

For example, $x \land (\lnot x \lor u) \le u$ does not hold in $O_6$ of~\autoref{fig:O6M4} (take $x \mapsto b, \ u \mapsto a$ as a counterexample). On the other hand, in any ortholattice, if $x \land (\lnot x \lor u) = 1$ then $u=1$. Indeed, consider any ortholattice, and suppose $x \land (\lnot x \lor u) = 1$. Recall that, in any bounded lattice with $1$ as a top element, if $p \land q = 1$ then $p=1$ and $q=1$ because $1 \le p \land q \le p$. 
In our example, we conclude $x = 1$ and $(\lnot x \lor u) = 1$. Now, substituting $x=1$ and using $\lnot 1 = 0$ gives us $u=1$. 

Such algebraic reasoning has a counterpart in proof-theoretic derivations.
We present in Section~\ref{sec:proofsystem} a system for derivation of formulas from axioms. Our system is complete for algebraic reasoning in ortholattices, allowing us to derive $u$ if we allow $x \land (\lnot x \lor u)$ as an axiom. Importantly, proof search in our system remains polynomial time, a result that we establish by showing a generalized notion of cut elimination.
%In this system, we do not have a deduction rule and modus ponens, which is what makes the access to axioms particularly important: we cannot equivalently rewrite a formula as an implication.

The use of axioms (equivalently, ortholattice presentations) cannot emulate \emph{all} instances of classical propositional logic axioms (indeed, proof search in our system remains in polynomial time instead of coNP). However, the above example hints that we can indeed use axioms to prove a larger set of classically valid problems than by using one monolithic formula in orthologic. Indeed, we show a number of practically important classes of problems for which reasoning in orthologic from axioms is complete, pointing to scenarios where orthologic may find useful applications.

\subsection{Contributions}

This paper shows how to use ortholattice reasoning with axioms as a sound polynomial-time deductive approach. We make the following contributions:
\begin{itemize} % break them apart if needed and introduce sub-bullets
    \item We first show that a proof system for orthologic with axioms satisfies a form of the Cut Elimination property, where Cut rules are restricted to eliminate only axioms and can only appear near leaves in the proof. From this, we deduce a subformula property.
    
    \item We show that, in the presence of axioms, there is an orthologic backward proof search procedure with worst case asymptotic time $\mathcal O(n^2(|A|+1))$, where $n$ is the size of the problem and $|A|$ the total number of axioms. Without axioms, the algorithm is quadratic.
    
    \item We study how orthologic with axioms can help solve some classes of classical problems and show that special case of satisfiability instances (2CNF, Horn clauses, renamed Horn, q-Horn, extended Horn) admit orthologic proofs, i.e. these problems are satisfiable in orthologic if and only if they are satisfiable in classical logic. 
    
    \item We show that orthologic decision problems can be flattened, similarly to the Tseitin transform \citep{tseitinComplexityDerivationPropositional1983}, and we use this to give an upper bound on the proving power of orthologic in terms of the width of classical resolution proofs.
    
    \item We show how orthologic reasoning can be extended to fragments of \emph{predicate} logic. We show that such quantified orthologic agrees with classical logic on the semantic of Datalog, and hence that Datalog programs admit $\OL$ proofs, making $\OL$ another possible generalization of Datalog to logic programming with negation and disjunction.
    
%    \item Finally, we show how orthologic can be extended to \textit{predicate} orthologic, a generalization of orthologic to quantified formulas which is sound with respect to first order logic. We extend the proof system with quantifiers and show Cut elimination.
\end{itemize}
Note that the above-mentioned consequences for Horn-like problems, Tseitin's transform, and Datalog extensions all crucially rely on being able to solve problems that consist of both axioms and a formula to prove from them; they cannot be reduced to proving an unconditional validity of an orthologic formula.

\section{Preliminaries}
We briefly present key concepts and notation which will be used in the present article. Ortholattices are the algebraic variety whose equational laws are presented in \autoref{tab:algebraiclaws}. As in any lattice, we can define an order relation $\leq$ by:
$$a \leq b  \iff (a=(a\land b))$$
which is also equivalent to $(b=(a\lor b))$.
This yields a partially ordered set (poset) whose corresponding axiomatization is shown in \autoref{tab:posetlaws} \cite{kalmbachOrthomodularLattices1983, meinanderSolutionUniformWord2010}. Note also that for any terms $x, y$ we have $x=y$ if and only if both $x\leq y$ and $y\leq x$. The relation $=$ defined this way is a congruence relation for $\leq$, $\land$, $\lor$ and $\neg$ and thus becomes equality in the quotient structure. From the point of view of first-order logic with equality, 
each model of Table~\ref{tab:algebraiclaws} axioms can be extended to a model of Table~\ref{tab:posetlaws} by defining the inequality $a \le b$ as the truth value of the atomic formula $a=(a\land b)$. Moreover, we have the converse: each model of Table~\ref{tab:posetlaws} axioms induces a quotient structure with respect to the $x \le y\ \&\ y \le x$ relation; this structure is a model of Table~\ref{tab:algebraiclaws} axioms.
\begin{table}[bth]
    \centering
    \begin{tabular}{r c @{\hskip 2em} | @{\hskip 2em} r c}
         P1: & $x \leq x$\\
         P2: & $x\leq y \hspace{0.5em} \& \hspace{0.5em} y \leq z \implies x \leq z$\\
         P3: & $0 \leq x$ & P3': & $x \leq 1$\\
         P4: & $x \land y \leq x$ & P4': & $x \leq x \lor y$\\
         P5: & $x \land y \leq y$ & P5': & $y \leq x \lor y$\\
        
         P6: & $x \leq y \hspace{0.5em} \& \hspace{0.5em} x \leq z \implies x \leq y\land z$ & 
            P6': & $x \leq z \hspace{0.5em} \& \hspace{0.5em} y \leq z \implies x\lor y \leq z$\\
         P7: & $x \leq \neg \neg x$ & P7': & $\neg \neg x \leq x$\\
         P8: & $x \leq y \implies \neg y \leq \neg x$\\
         P9: & $x \land \neg x \leq 0$ & P9': & $1 \leq x \lor \neg x$\\
    \end{tabular}
    
    \
    
    \caption{Axiomatization of ortholattices in the signature $(S, \leq, \land, \lor, 0, 1, \neg)$ as partially ordered sets. We use $\&$ for conjunction between atomic formulas of axioms, to differentiate it from the term-level lattice operation $\land$.
    This axiomatization corresponds to one in Table~\ref{tab:algebraiclaws}: in one direction, define $x \le y$ to be $x \land y = x$; in the other direction, define $x=y$ to be $(x \le y)\ \&\ (y \le x)$.
    \label{tab:posetlaws}}
\end{table}

By definition, Boolean algebras are precisely ortholattices that are distributive. \noindent\autoref{fig:O6M4} shows ortholattices $O_6$ and $M_4$ that are \emph{not} Boolean algebras. In fact, an ortholattice is a Boolean algebra if and only if it does not contain $O_6$ nor $M_4$ as a sub-ortholattice
\cite{daveyIntroductionLatticesOrder2002, kalmbachOrthomodularLattices1983}.
Despite being strictly weaker than laws of Boolean algebra, properties in~\autoref{tab:algebraiclaws} and~\autoref{tab:posetlaws} allow us to prove a number of desirable facts. This includes all laws of bounded lattices (absorption, reordering and de-duplicating conjuncts and disjuncts), as well as laws relating complement to lattice operators (such as the laws needed to transform formulas to negation-normal form). Laws of Boolean algebra that do not necessarily hold include distributivity, modularity, and properties such as $(\lnot x \lor y) = 1$ implying $x \le y$.

Similarly to how Boolean Algebra is the algebraic structure corresponding to classical logic, 
%and how Heyting algebras serves this purpose for intuitionistic logic, 
ortholattices form a class of structures that corresponds to a logic, \textit{orthologic}, for which we study proof-theoretic and algorithmic properties in the following sections. We denote classical logic by $\CL$ and orthologic by $\OL$.
\begin{defin}[Terms]
$\mathcal{T}_\OL$ denotes the term algebra over the signature of ortholattices over a fixed countably infinite set of variables, that is, the set of all terms which can be built from variables and $(\land, \lor, 0, 1, \neg)$. We typically represent terms with lowercase Greek letters and variables with $x,y,z$, possibly with indices. Terms are constructed inductively as trees. Leaves are labeled with $0$, $1$, or variables. Nodes are labeled with logical symbols. Since $\lor$ and $\land$ are commutative, the children of a node form a set (non-ordered).
$\mathcal{T}_\CL = \mathcal{T}_\OL$, and is also the set of formulas for both classical logic and orthologic.

\end{defin}

Note that the laws of both $\OL$ and $\CL$ imply that $0$ can always be represented by $x \land \neg x$ and $1$ as $x \lor \neg x$. To simplify proofs, we thus may omit the cases corresponding to $0$ and $1$.

The \textit{word problem} for an algebra consists in, given two terms in the language of the algebra, deciding if they are always equal by the laws of the algebra or not. For ortholattices, we can relax the definition to allow inequality queries, as they can be expressed as equivalent equalities.
A (finite) \textit{presentation} for an algebra is a (finite) set of equalities (which we relax to inequalities in ortholattices) $\lbrace \phi_1 \leq \psi_1, ..., \phi_n\leq \psi_n \rbrace$. The \textit{uniform word problem} for presented ortholattices is the task consisting in, given a presentation $A$ and two terms $\phi$ and $\psi$, deciding if $\phi\leq \psi$ follows from the laws of ortholattices and the axioms in $A$. 

In the terminology of classical first-order logic, the laws of ortholattice in \autoref{tab:posetlaws} are a finite set of \emph{universally quantified} formulas, $\mathcal{T}$, and they define a first-order theory. The presentation (axioms) $A$ is a set of \emph{quantifier-free} formulas, whereas $\phi\leq \psi$ is also a quantifier-free formula, with variables possibly in common with those of $A$. We can then view the uniform word problem as a special case of the question of semantic consequence in first-order logic:
$
    \mathcal{T} \cup A\ \models\ \phi \le \psi
$.

\section{Complete Proof System and Cut Elimination}
\label{sec:proofsystem}

We formulate our proof system for orthologic as a sequent calculus. We represent sequents by decorating the formulas with superscript ${}^L$ or ${}^R$, depending on whether they appear on the left or right side. For example, $\phi^L, \phi^R$ stands for $\phi \vdash \psi$ in more conventional notation.
\begin{defin}
    If $\phi$ is a formula, we call $\phi^L$ and $\phi^R$ annotated formulas.
    A \textit{sequent} is a set  of at most two annotated formulas. We use $\Gamma$ and $\Delta$ to represent sets that are either empty or contain exactly one annotated formula ($|\Gamma| \le 1, |\Delta| \le 1$).
\end{defin}
\autoref{fig:proofSystem} shows our sequent calculus for orthologic, parameterised by a set of sequents $A$ called axioms. In the present article, \textit{orthologic}, or $\OL$ denotes this specific proof system. Without the support for arbitrary axioms, an equivalent system, with a different presentation, was introduced in \cite{schultemontingCutEliminationWord1981}.

Note that the axioms we consider in this section are not universally quantified: they refer to arbitrary but fixed propositions.

\begin{figure}[ht]
    \begin{subfigure}{0.7\textwidth}
        \begin{tabular}{l l}
        \ \\
            \multicolumn{2}{c}{            
                \AxiomC{}
                \RightLabel{\text { Hyp}}
                \UnaryInfC{$\phi^L, \phi^R$}
                \DisplayProof
            }\\[3ex]

            \multicolumn{2}{c}{
                \AxiomC{$\Gamma, \psi^R$}
                \AxiomC{$\psi^L, \Delta$}
                \RightLabel{\text{ Cut}}
                \BinaryInfC{$\Gamma, \Delta$}
                \DisplayProof
            }\\[3ex]

            \multicolumn{2}{c}{
                \AxiomC{$\Gamma$}
                \RightLabel{\text { Weaken}}
                \UnaryInfC{$\Gamma, \Delta$}
                \DisplayProof
            }\\[3ex]

            \AxiomC{$\Gamma, \phi^L$}
            \RightLabel{\text { LeftAnd}}
            \UnaryInfC{$\Gamma, (\phi \land \psi)^L$}
            \DisplayProof &
            \AxiomC{$\Gamma, \phi^R$}
            \AxiomC{$\Gamma, \psi^R$}
            \RightLabel{\text{ RightAnd}}
            \BinaryInfC{$\Gamma, (\phi \land \psi)^R$}
            \DisplayProof
            \\[3ex]

            \AxiomC{$\Gamma, \phi^L$}
            \AxiomC{$\Gamma, \psi^L$}
            \RightLabel{\text{ LeftOr}}
            \BinaryInfC{$\Gamma, (\phi \lor \psi)^L$}
            \DisplayProof &
            \AxiomC{$\Gamma, \phi^R$}
            \RightLabel{\text{ RightOr}}
            \UnaryInfC{$\Gamma, (\phi \lor \psi)^R$}
            \DisplayProof
            \\[3ex]

            \AxiomC{$\Gamma, \phi^R$}
            \RightLabel{\text { LeftNot}}
            \UnaryInfC{$\Gamma, (\neg \phi)^L$}
            \DisplayProof &
            \AxiomC{$\Gamma, \phi^L$}
            \RightLabel{\text{ RightNot}}
            \UnaryInfC{$\Gamma, (\neg \phi)^R$}
            \DisplayProof \\
\ \\

        \end{tabular}
        \caption{Deduction rules of Orthologic. Each holds for arbitrary $\Gamma$, $\Delta$, $\phi$, $\psi$}
    \end{subfigure}
\\[2em]
    \begin{subfigure}{0.7\textwidth}
        \begin{center}
            \AxiomC{}
            \RightLabel{\text{ Ax($\Gamma$,$\Delta$) \hspace{1em} If \ $\Gamma \cup \Delta\ \in A$}}
            \UnaryInfC{$\Gamma, \Delta$}
            \DisplayProof 
        \end{center}

        \caption{Rule for additional non-logical axioms.}
    \end{subfigure}

%}
\caption{Sequent-calculus rules for Orthologic derivations from a set of axioms $A$. Sets $\Gamma$ and $\Delta$ are either empty or a single formula on any side.}
\label{fig:proofSystem}

\end{figure}
One can think of this proof system as Gentzen's sequent calculus for classical logic~\cite{gentzenUntersuchungenUberLogische1935} restricted to ensure the following syntactic invariant:
\begin{center}\emph{
At any given place in a proof, a sequent never has more than two formulas on both sides combined.}
\end{center}
This restriction on the proof system bears resemblance to the syntactic restriction of intuitionistic sequent calculus, where a sequent can never have more than one formula on the right side, a restriction lifted in the classical logic sequent calculus system. Compared to intuitionistic logic, orthologic allows us to prove $\vdash \phi, \lnot \phi$, represented as $\phi^R, (\neg \phi)^R$, using the following steps.
\begin{center}
  \AxiomC{}
  \RightLabel{\text { Hyp}}
  \UnaryInfC{$\phi^L, \phi^R$}
  \RightLabel{\text{ RightNot}}
  \UnaryInfC{$\phi^R, (\neg \phi)^R$}
  % add or not?
%  \RightLabel{\text{ RightOr}}
%  \UnaryInfC{$\phi^R, (\neg \phi \lor \phi)^R$}
%  \RightLabel{\text{ RightOr}}
%  \UnaryInfC{$(\phi \lor \neg \phi)^R, (\neg \phi \lor \phi)^R$}
  
  \DisplayProof
\end{center}
On the other hand, orthologic restricts the number of assumptions on the left side of the sequent. This strong restriction will be rewarded by the existence of a polynomial-time proof search procedure.
\begin{defin} \label{defin:admissible}
 We say that a deduction rule is \textit{admissible} if any sequent that can be proven with the rule can be proven without. 
\end{defin}

\subsection{Ortholattice Semantics for Orthologic}

We interpret a sequent $\phi^L, \psi^R$ as $\phi \le \psi$ in an ortholattice. More generally, we have the following definition.
\begin{defin}
\label{defin:interpretation} The \emph{interpretation} of a sequent is given by the following table, where $\emptyset$ denotes the empty sequent:
\[\renewcommand{\arraystretch}{1.2}
\begin{array}{c|c}
\mbox{sequent $S$}     & \mbox{ortholattice inequality $\overline{S}$ (up to equivalence)} \\ \hline\hline
\phi^L, \psi^R     & \phi \leq \psi  \\ \hline
\phi^L, \psi^L     & \phi \leq \neg\psi \\ \hline
\phi^R, \psi^R     & \neg\phi \leq \psi \\ \hline
\phi^L             & \phi \leq 0 \\ \hline
\phi^R             & 1 \leq \phi \\ \hline
\emptyset          & 1 \leq 0
\end{array}
\]
The intended reading of the table above is a mapping of sequents (which are sets) to ortholattice atomic formulas \emph{up to logical equivalence}. The set
$\{\phi^L, \psi^L\}$ can be mapped to either $\phi \leq \neg\psi$ or to
$\neg\psi \le \phi$, but these are equivalent in an ortholattice (analogously for mapping $\{\phi^R, \psi^R\}$).

The interpretation of a deduction rule in~\autoref{fig:proofSystem} with $k$ premises $P_1,\ldots, P_k$ %where $k \in \{ 0, 1, 2 \}$ 
and a conclusion $C$, is the universally quantified first-order logic formula
$
    \overline{P}_1 \ \& \  \ldots \ \& \  \overline{P}_n \implies \overline{C}
$.
\end{defin}
Given an axiom set $A$ we talk about ortholattice with presentation $A$ by taking the interpretation of all axioms in $A$.
Our proof system can prove every axiom of ortholattices (\autoref{tab:posetlaws}) and, conversely.
\begin{lemma}[Soundness and Completeness of $\OL$ Deduction Rules]
\label{lem:soundCompPropOl}
  %The first-order theory of the translations of rules in orthologic and axioms in $A$ (\autoref{fig:proofSystem}) is equivalent to the first-order theory of ortholattices (\autoref{tab:posetlaws}) with presentation $A$.\\
Let $A$ be an arbitrary (possibly infinite) set of axioms.
  A sequent has a derivation from $A$ using the rules of orthologic (\autoref{fig:proofSystem}) iff its interpretation is in the first-order theory of ortholattices (\autoref{tab:posetlaws}) with presentation $A$.
\end{lemma}
\begin{proof}
    Sketch. For every \autoref{tab:posetlaws} law of the form $\overline{P}_1 \land \ldots \land \overline{P}_n \rightarrow \overline{C}$, a matching deduction rule 
    \begin{center}
        \AxiomC{$P_1$}
        \AxiomC{$...$}
        \AxiomC{$P_n$}
        \TrinaryInfC{$C$}
        \DisplayProof
    \end{center}
    is easily seen to be admissible. Conversely, for every deduction rule of \autoref{fig:proofSystem}, the corresponding law is a consequence (in first order logic) of the axioms of \autoref{tab:posetlaws}.
\end{proof}

For any axiom set $A$, this makes our system (with the Cut rule) sound and complete for the class of all ortholattices satisfying axioms in $A$. Note that this interpretation is compatible with the interpretation of sequents in classical logic.
We can use the soundness and completeness to obtain simple model-theoretic proofs for orthologic.
We can show, for example, that substitution of equivalent formulas is admissible.
\begin{lemma}
\label{lem:substformulas}
    Let $\Gamma$ and $\Delta$ denote sets with at most one labelled formula each. Let
    $\Gamma[\chi:=\phi]$ denote the substitution inside $\Gamma$ of $\chi$ (a placeholder formula symbol) by $\psi$. The following rule for substitution of equivalent formulas is admissible in orthologic:
\begin{center}
    \AxiomC{$\Gamma[\chi:=\phi], \Delta[\chi:=\phi]$}
    \AxiomC{$\phi^L, \psi^R$}
    \AxiomC{$\psi^L, \phi^R$}
    \RightLabel{\textup{ Subst}}
    \TrinaryInfC{$\Gamma[\chi:=\psi], \Delta[\chi:=\psi]$}
    \DisplayProof 
\end{center}
Said otherwise, if both $\phi^L, \psi^R$ and $\psi^L, \phi^R$ can be shown then
arbitrary occurrences of $\phi$ in a proven sequent can be replaced by $\psi$. 
\end{lemma}
\begin{proof}
    The argument is semantic. Fix any ortholattice $\mathcal O$ satisfying the axioms.
    Since both ${\phi^L, \psi^R}$ and $\psi^L, \phi^R$ are provable, it follows that $\phi = \psi$ in $\mathcal O$. Hence, 
    ${\Gamma[\chi:=\phi] = \Gamma[\chi:=\psi]}$ and ${\Delta[\chi:=\phi] = \Delta[\chi:=\psi]}$. By completeness, the sequent ${\Gamma[\chi:=\psi], \Delta[\chi:=\psi]}$ is provable.
\end{proof}

\subsection{Partial Cut Elimination}

As a sequent calculus, our system has structural rules, introduction rules for each logical symbol and a Cut rule, but no elimination rule. Consequently, by inspecting all rules, we conclude that \emph{Cut is the only rule whose premises contain formulas that are not subformulas of the concluding sequent}.
\begin{defin} \label{defin:cutAndPrincipalFormulas}
    In an instance of a Cut rule in \autoref{fig:proofSystem}, we call the formula $\psi$ the \textit{cut formula}. In an instance of a left or right rule, the newly constructed formula is called the \textit{principal formula}. In the Weaken rule, if $\Delta$ contains a formula, it is also called the principal formula.
\end{defin}

\citep{schultemontingCutEliminationWord1981} showed that orthologic, without arbitrary non-logical axioms, admits cut elimination. 
The crucial challenge is that, in contrast to classical or intuitionistic calculus, we cannot simply add additional assumptions to the left-hand side of sequents in orthologic derivations. The reason is the restriction on the number of formulas in sequents. The following example illustrates this phenomenon.
\begin{example}
    We saw in \autoref{ex:intro} that $(x \land (\neg x \lor u)) \leq u$ is not always valid. In particular, the sequent $(x \land (\neg x \lor u))^L, u^R$ is not provable in orthologic without axioms.
    However, with axiom $(x \land (\neg x \lor u))^R$, the sequent $u^R$ is provable as follows. First, $(\neg x \lor u)^R$ is provable:
    \begin{center}  
        \AxiomC{}
        \RightLabel{Ax}
        \UnaryInfC{$(x \land (\neg x \lor u))^R$}

        \AxiomC{}
        \RightLabel{ \hypname}
        \UnaryInfC{$(\neg x \lor u)^L, (\neg x \lor u)^R$}
        \RightLabel{L.And}
        \UnaryInfC{$(x \land (\neg x \lor u))^L, (\neg x \lor u)^R$}
        \RightLabel{ Cut}
        \BinaryInfC{$(\neg x \lor u)^R$}

        \DisplayProof
    \end{center}
    Then, 
    \begin{center} 

        \AxiomC{$(\neg x \lor u)^R$}

        \AxiomC{}
        \RightLabel{Ax}
        \UnaryInfC{$(x \land (\neg x \lor u))^R$}
        \AxiomC{}
        \RightLabel{ \hypname}
        \UnaryInfC{$x^L, x^R$}
        \RightLabel{L.And}
        \UnaryInfC{$(x \land (\neg x \lor u))^L, x^R$}

        \RightLabel{ Cut}
        \BinaryInfC{$x^R$}
        
        \RightLabel{ L.Not}
        \UnaryInfC{$(\neg x)^L$}
        \RightLabel{ Weaken}
        \UnaryInfC{$(\neg x)^L, u^R$}

        \AxiomC{}
        \RightLabel{\hypname}
        \UnaryInfC{$u^L, u^R$}
        \RightLabel{ L.Or}
        \BinaryInfC{$(\neg x \lor u)^L, u^R$}

        \RightLabel{ Cut}
        \BinaryInfC{$u^R$}
        
        \DisplayProof
    \end{center}
    In a \emph{classical} sequent calculus system, the above derivation using the axiom $x \land (\lnot x \lor u)$ could be transformed into a new derivation where each sequent has an additional assumption $x \land (\lnot x \lor u)$ and where the use of axiom rule is replaced with the use of the Hyp rule. This transformation does not apply to $\OL$: it would create sequents with more than two formulas, which, by the definition in \autoref{fig:proofSystem} cannot appear in $\OL$ proofs.
\end{example}
For this reason, the ability to add non-logical axioms is crucial in orthologic. We aim to extend the cut elimination property to proofs containing arbitrary axioms.  This will allow us to devise an efficient decision procedure for orthologic with axioms, and, by extension, the word problem for finitely presented ortholattices. Moreover, the proof we present is constructive, in the sense that it shows an algorithmic way to eliminate instances of the Cut rule from a proof. Furthermore, we need not worry about the size of the transformed proof, because our Cut elimination property will enable us to derive a subformula property and a bound on the size of the proof of any given formula.

However, the system does \textit{not} allow for complete cut elimination in the presence of axioms, as the following short example shows.
\begin{example}
    Let $x_1,x_2,y$ be distrinct variables and let the sequent $(x_1 \lor x_2)^L, y^R$ be the only axiom. The sequent $x_1^L, y^R$ is then provable:
    \vspace{0.3em}
    \begin{center}
        \AxiomC{}
        \RightLabel{\textup{ \hypname}}
        \UnaryInfC{$x_1^L, x_1^R$}
        \RightLabel{\textup{ RightOr}}
        \UnaryInfC{$x_1^L, (x_1 \lor x_2)^R$}
        \AxiomC{}
        \RightLabel{\textup{ Ax}}
        \UnaryInfC{$(x_1 \lor x_2)^L, y^R$}
        \BinaryInfC{$x_1^L, y^R$}
        \DisplayProof
    \end{center}    
    but it cannot be proven without using the Cut rule. To see why, note that Hyp, LeftAnd, RightAnd, LeftOr, RightOr, LeftNot, RightNot do not yield sequents whose syntactic form can be $x_1^L, y^R$. Furthermore, Ax does not produce the desired sequent as it is not the axiom. Finally, Weaken does not help because its premise would not be provable: neither $x_1^L$ nor $y^R$ are individually provable from the axiom, as can be seen by a semantic argument: it could be, for example, that both $x_1$ and $y$ have value $0$ or both have value $1$ in an ortholattice that satisfies the axiom.
\end{example}
\noindent Thus, cut rule is in general necessary when reasoning from axioms, and we need to formulate a suitable generalization of the concept of cut elimination.
For this purpose, we define the \textit{rank} of an instance of the cut rule.
\begin{defin}
    An instance of the Cut rule has \textit{rank 1} if either of its premises is an axiom. It has \textit{rank 2} if either of its premises is the conclusion of a rank 1 Cut rule. 
\end{defin}
The following theorem is our main result. It implies that the Cut rule can be eliminated or restricted to only cut with respect to axioms. Part (1) has immediate consequences for the subformula property. Part (2) gives further insight into normalized proofs, further restricts  our proof procedure in the next section, and it helps with the inductive argument in the proof of the theorem.
\begin{thm}[Cut Elimination for Orthologic]
\label{thm:cutelim}
If a sequent is provable in the system of \autoref{fig:proofSystem} with axioms

\begin{center}
\vspace{-0.3em}
    \AxiomC{}
    \RightLabel{\text{ Ax($a_i^\circ$,$b_i^\Box$)}}
    \UnaryInfC{$a_i^\circ, b_i^\Box$}
    \DisplayProof

\end{center}
for all $(a_i^\circ$,$b_i^\Box) \in A$ ($a_i$ and $b_i$ are formulas, $^\circ$ and $^\Box$ are side annotation), then there is a proof of that sequent from the same axioms such that:
\begin{enumerate}
    \item All instances of the Cut rule use only formulas among  $a_1, ... a_n, b_1,..., b_n$ as cut formulas.
    \item All instances of the Cut rule are rank 1 or 2.
    %\item All formulas in the conclusion of cut rules contain only subformulas of $x_1, ... x_n, y_1,..., y_n$ (or some variant of that property).
\end{enumerate}

\end{thm}

\begin{proof}
If a proof does not satisfy the properties (1) and (2) of the theorem statement, then
there is a Cut rule for which the condition in (1) or (2) does not hold. 
Consider a derivation using such a Cut rule as the last step; when Cut is not bottommost, the properties follow trivially by induction. Let the proof be of the form:
\begin{center}
    \AxiomC{$\A$}
    \UnaryInfC{$\Gamma, \psi^R$}
    \AxiomC{$\B$}
    \UnaryInfC{$\psi^L, \Delta$}
    \RightLabel{\text{ Cut}}
    \BinaryInfC{$\Gamma, \Delta$}
    \DisplayProof
\end{center}
where $\A$ and $\B$ are the proof trees whose conclusions are respectively the left and right premises and $\psi$ is the Cut formula. We show that there exists a proof of $\Gamma, \Delta$ that satisfies the two properties of the theorem. 

We proceed by induction on the length of the proof% and the size of the cut formula
. Hence, we can assume by induction that $\A$ and $\B$ satisfy properties 1 and 2. By induction hypothesis, it suffices to show how to transform the proof of $\Gamma, \Delta$ into one where Cut is used only in subproofs strictly smaller than the proof we started with.
We do case analysis on $\A$ and $\B$, showing for each case how to transform the proof in this way. $\hookrightarrow$ denotes this transformation.

\paragraph{Case 1} Suppose $\A$ ends (and hence starts) with a Hypothesis rule. Then, $\Gamma = \psi^L$ and $\psi^L, \Delta$ can be reached using only $\B$. The case where $\B$ is a Hypothesis rule is symmetric.
\paragraph{Case 2} ($\A$ ends with Weaken)
\subparagraph{Case 2.a} Suppose $\A$ ends with a Weaken rule and $\psi^R$ is not the \emph{principal formula} (see \autoref{defin:cutAndPrincipalFormulas}).
\begin{center}
\begin{tabular}{c c c}
    \AxiomC{$\A'$}
    \UnaryInfC{$\psi^R$}
    \RightLabel{\text { Weaken}}
    \UnaryInfC{$\Gamma, \psi^R$}
    \AxiomC{$\B$}
    \UnaryInfC{$\psi^L, \Delta$}
    \RightLabel{\text{ Cut}}
    \BinaryInfC{$\Gamma, \Delta$}
    \DisplayProof &
    
    $\hookrightarrow$ &
    
    \AxiomC{$\A'$}
    \UnaryInfC{$\psi^R$}
    \AxiomC{$\B$}
    \UnaryInfC{$\psi^L, \Delta$}
    \RightLabel{\text{ Cut}}
    \BinaryInfC{$\Delta$}
    \RightLabel{\text { Weaken}}
    \UnaryInfC{$\Gamma, \Delta$}
    
    \DisplayProof
\end{tabular}
\end{center}
In the transformed proof, $\A'$ is part of $\A$, so Cut applies to a smaller subproof  and can be transformed to satisfy the properties (1) and (2) by inductive hypothesis.

\subparagraph{Case 2.b} Suppose $\A$ ends with a Weaken rule and $\psi^R$ is the principal formula.
\begin{center}
\begin{tabular}{c c c}
    \AxiomC{$\A'$}
    \UnaryInfC{$\Gamma$}
    \RightLabel{\text { Weaken}}
    \UnaryInfC{$\Gamma, \psi^R$}
    \AxiomC{$\B$}
    \UnaryInfC{$\psi^L, \Delta$}
    \RightLabel{\text{ Cut}}
    \BinaryInfC{$\Gamma, \Delta$}
    \DisplayProof &
    
    $\hookrightarrow$ &
    
    \AxiomC{$\A'$}
    \UnaryInfC{$\Gamma$}
    \RightLabel{\text { Weaken}}
    \UnaryInfC{$\Gamma, \Delta$}
    
    \DisplayProof
    \\[5ex]
\end{tabular}
\end{center}

\paragraph{Case 3} ($\A$ ends with a Left rule where $\psi$ is not principal)
\subparagraph{Case 3.a} Suppose $\A$ ends with a LeftAnd rule where $\Gamma = (\alpha \land \beta)^L$
\begin{center}
\begin{tabular}{c c c}
    \AxiomC{$\A'$}
    \UnaryInfC{$\alpha^L, \psi^R$}
    \RightLabel{\text { LeftAnd}}
    \UnaryInfC{$(\alpha \land \beta)^L, \psi^R$}
    \AxiomC{$\B$}
    \UnaryInfC{$\psi^L, \Delta$}
    \RightLabel{\text{ Cut}}
    \BinaryInfC{$(\alpha \land \beta)^L, \Delta$}
    \DisplayProof &
    
    $\hookrightarrow$ &
    
    \AxiomC{$\A'$}
    \UnaryInfC{$\alpha^L, \psi^R$}
    \AxiomC{$\B$}
    \UnaryInfC{$\psi^L, \Delta$}
    \RightLabel{\text{ Cut}}
    \BinaryInfC{$\alpha^L, \Delta$}
    \RightLabel{\text { LeftAnd}}
    \UnaryInfC{$(\alpha \land \beta)^L, \Delta$}
    \DisplayProof
    \\[5ex]
\end{tabular}
\end{center}

\subparagraph{Case 3.b} Suppose $\A$ ends with a LeftOr rule where $\phi = \alpha \lor \beta$
\begin{center}
\begin{tabular}{c c c}
    \AxiomC{$\A'$}
    \UnaryInfC{$\alpha^L, \psi^R$}
    \AxiomC{$\A''$}
    \UnaryInfC{$\beta^L, \psi^R$}
    \RightLabel{\text { LeftOr}}
    \BinaryInfC{$(\alpha \lor \beta)^L, \psi^R$}
    \AxiomC{$\B$}
    \UnaryInfC{$\psi^L, \Delta$}
    \RightLabel{\text{ Cut}}
    \BinaryInfC{$(\alpha \lor \beta)^L, \Delta$}
    \DisplayProof \\[6ex]
    
    $\hookrightarrow$ \\
    
    \AxiomC{$\A'$}
    \UnaryInfC{$\alpha^L, \psi^R$}
    \AxiomC{$\B$}
    \UnaryInfC{$\psi^L, \Delta$}
    \RightLabel{\text{ Cut}}
    \BinaryInfC{$\alpha ^L, \Delta$}
    \AxiomC{$\A''$}
    \UnaryInfC{$\beta^L, \psi^R$}
    \AxiomC{$\B$}
    \UnaryInfC{$\psi^L, \Delta$}
    \RightLabel{\text{ Cut}}
    \BinaryInfC{$\beta^L, \Delta$}
    \RightLabel{\text { LeftOr}}
    \BinaryInfC{$(\alpha \lor \beta)^L, \Delta$}
    \DisplayProof
    \\[2ex]
\end{tabular}
\end{center}

\subparagraph{Case 3.c} Suppose $\A$ ends with a LeftNot rule, i.e. $\Gamma = \neg \alpha$
\begin{center}
\begin{tabular}{c c c}
    \AxiomC{$\A'$}
    \UnaryInfC{$\alpha^R, \psi^R$}
    \RightLabel{\text { LeftNot}}
    \UnaryInfC{$(\neg \alpha)^L, \psi^R$}
    \AxiomC{$\B$}
    \UnaryInfC{$\psi^L, \Delta$}
    \RightLabel{\text{ Cut}}
    \BinaryInfC{$\alpha^L, \Delta$}
    \DisplayProof &
    
    $\hookrightarrow$ &
    
    \AxiomC{$\A'$}
    \UnaryInfC{$\alpha^R, \psi^R$}
    \AxiomC{$\B$}
    \UnaryInfC{$\psi^L, \Delta$}
    \RightLabel{\text { Cut}}
    \BinaryInfC{$\alpha^R, \Delta$}
    \RightLabel{\text { LeftNot}}in
    \UnaryInfC{$(\neg \alpha)^L, \Delta$}
    \DisplayProof
    \\[2ex]
\end{tabular}
\end{center}

The cases where $\B$ ends with a Right rule are symmetric. 
\paragraph{Case 4} If $\A$ ends with Right rule where $\psi^R$ is not the principal formula, the transformation is symmetric to the Left rule case. Similarly if $\B$ ends with a Left rule where $\psi^L$ is not the principal formula.

\paragraph{Case 5} ($\A$ ends with Right rule and $\B$ with a Left rule, $\psi$ is principal in both)

\subparagraph{Case 5.a} Suppose $\A$ ends with a RightOr rule, i.e. $\psi = (\alpha \lor \beta)$. In this case, $\B$ has to end with a LeftOr rule, as it is the only Left rule that can produce $(\alpha \lor \beta)^L$
\begin{center}
\begin{tabular}{c c c}
    \AxiomC{$\A'$}
    \UnaryInfC{$\Gamma, \alpha^R$}
    \RightLabel{\text { RightOr}}
    \UnaryInfC{$\Gamma, (\alpha \lor \beta)^R$}
    \AxiomC{$\B'$}
    \UnaryInfC{$\alpha^L, \Delta$}
    \AxiomC{$\B''$}
    \UnaryInfC{$\beta^L, \Delta$}
    \RightLabel{\text { LeftOr}}
    \BinaryInfC{$(\alpha \lor \beta)^L, \Delta$}
    \RightLabel{\text{ Cut}}
    \BinaryInfC{$\Gamma, \Delta$}
    \DisplayProof &
    
    $\hookrightarrow$ &
    
    \AxiomC{$\A'$}
    \UnaryInfC{$\Gamma, \alpha^R$}
    \AxiomC{$\B'$}
    \UnaryInfC{$\alpha^L, \Delta$}
    \RightLabel{\text{ Cut}}
    \BinaryInfC{$\Gamma, \Delta$}
    \DisplayProof
    \\[2ex]
\end{tabular}
\end{center}

\subparagraph{Case 5.b} If $\A$ ends with a RightAnd rule and $\B$ with a LeftAnd, the transformation is symmetric to case 5.a

\subparagraph{Case 5.c} If $\A$ ends with a RightNot rule and $\B$ a LeftNot rule, and $\phi = \neg \alpha$ is principal in both:
\begin{center}
\begin{tabular}{c c c}
    \AxiomC{$\A'$}
    \UnaryInfC{$\Gamma, \alpha^L$}
    \RightLabel{\text { RightNot}}
    \UnaryInfC{$\Gamma, (\neg\alpha)^R$}
    \AxiomC{$\B'$}
    \UnaryInfC{$\alpha^R, \Delta$}
    \RightLabel{\text { RightNot}}
    \UnaryInfC{$(\neg\alpha)^L, \Delta$}
    \RightLabel{\text{ Cut}}
    \BinaryInfC{$\Gamma, \Delta$}
    \DisplayProof &
    
    $\hookrightarrow$ &
    
    \AxiomC{$\A'$}
    \UnaryInfC{$\Gamma, \alpha^L$}
    \AxiomC{$\B'$}
    \UnaryInfC{$\alpha^R, \Delta$}
    \RightLabel{\text{ Cut}}
    \BinaryInfC{$\Gamma, \Delta$}
    \DisplayProof
    \\[2ex]
\end{tabular}
\end{center}

\paragraph{Case 6}
Suppose either $\A$ or $\B$ end with an Axiom rule. Then, properties 1 and 2 are immediate.

\paragraph{Case 7}
Suppose $\A$ ends with a Cut rule, which by induction we can assume is rank 1 or 2. 
%We are concerned with cases where B ends with a rule where the cut formula is principal  or a cut rule
\subparagraph{Case 7.a}
If $\A$ is rank 1, the following transformation works, since if $\A''$ is an axiom both properties immediately hold for both Cuts, and if $\A'$ is an axiom, both properties hold for the last cut and the other cut has smaller size.
\begin{center}
\begin{tabular}{c c c}
    \AxiomC{$\A'$}
    \UnaryInfC{$\Gamma, \phi^R$}
    \AxiomC{$\A''$}
    \UnaryInfC{$\phi^L, \psi^R$}
    \RightLabel{\text{ Cut}}
    \BinaryInfC{$\Gamma, \psi^R$}
    \AxiomC{$\B$}
    \UnaryInfC{$\psi^L, \Delta$}
    \RightLabel{\text{ Cut}}
    \BinaryInfC{$\Gamma, \Delta$}
    \DisplayProof &

    $\hookrightarrow$ &
 
    \AxiomC{$\A'$}
    \UnaryInfC{$\Gamma, \phi^R$}
    \AxiomC{$\A''$}
    \UnaryInfC{$\phi^L, \psi^R$}
    \AxiomC{$\B$}
    \UnaryInfC{$\psi^L, \Delta$}
    \RightLabel{\text{ Cut}}
    \BinaryInfC{$\phi^L, \Delta$}
    \RightLabel{\text{ Cut}}
    \BinaryInfC{$\Gamma, \Delta$}
    \DisplayProof\\[2ex]
\end{tabular}
\end{center}
\subparagraph{Case 7.b} Suppose $\A$ is rank 2 and it is $\A'$ that ends with a rank 1 Cut:
\begin{center}
\begin{tabular}{c c c}
    \AxiomC{$\A_1'$}
    \UnaryInfC{$\Gamma, \chi^R$}
    \AxiomC{$\A_2'$}
    \UnaryInfC{$\chi^L, \phi^R$}
    \RightLabel{\text{ Cut$^1$}}
    \BinaryInfC{$\Gamma, \phi^R$}
    \AxiomC{$\A''$}
    \UnaryInfC{$\phi^L, \psi^R$}
    \RightLabel{\text{ Cut$^2$}}
    \BinaryInfC{$\Gamma, \psi^R$}
    \AxiomC{$\B$}
    \UnaryInfC{$\psi^L, \Delta$}
    \RightLabel{\text{ Cut$^3$}}
    \BinaryInfC{$\Gamma, \Delta$}
    \DisplayProof\\[2ex]
\end{tabular}
\end{center}
Either $\A_1'$ or $\A_2'$ is an axiom. If $\A_2'$ is, the same transformation as above works because the last cut again has the axiom $\phi$ as a cut formula. If $\A_1'$ is an Axiom, we transform to the following:
\begin{center}
\begin{tabular}{c c c}

    $\hookrightarrow$ &

    \AxiomC{$\A_1'$}
    \UnaryInfC{$\Gamma, \chi^R$}
    \AxiomC{$\A_2'$}
    \UnaryInfC{$\chi^L, \phi^R$}
    \AxiomC{$\A''$}
    \UnaryInfC{$\phi^L, \psi^R$}\RightLabel{\text{ Cut}$^1$}
    \BinaryInfC{$\chi^L, \psi^R$}
    \AxiomC{$\B$}
    \UnaryInfC{$\psi^L, \Delta$}
    \RightLabel{\text{ Cut}$^2$}
    \BinaryInfC{$\chi^R, \Delta$}
    \RightLabel{\text{ Cut$^3$}}
    \BinaryInfC{$\Gamma, \Delta$}
    \DisplayProof\\[2ex]
\end{tabular}
\end{center}
Then, the new Cut$^3$ is of rank 1 and its cut formula is part of an axiom. Cut$^2$ is of strictly smaller size than the proof we started from, so by induction its conclusion can be obtained with a proof satisfying properties 1 and 2.
\subparagraph{Case 7.c}
Suppose now that $\A''$ is a rank 1 Cut. We transform the proof as follows:
\begin{center}
\begin{tabular}{c c c}

    \AxiomC{$\A'$}
    \UnaryInfC{$\Gamma, \phi^R$}    
    \AxiomC{$\A_1''$}
    \UnaryInfC{$\phi^L, \chi^R$}
    \AxiomC{$\A_2''$}
    \UnaryInfC{$\chi^L, \psi^R$}
    \RightLabel{\text{ Cut}}
    \BinaryInfC{$\phi^L, \psi^R$}
    \RightLabel{\text{ Cut}}
    \BinaryInfC{$\Gamma, \psi^R$}
    \AxiomC{$\B$}
    \UnaryInfC{$\psi^L, \Delta$}
    \RightLabel{\text{ Cut}}
    \BinaryInfC{$\Gamma, \Delta$}
    \DisplayProof\\[6ex]
    
    $\hookrightarrow$ \\

    \AxiomC{$\A'$}
    \UnaryInfC{$\Gamma, \phi^R$}    
    \AxiomC{$\A_1''$}
    \UnaryInfC{$\phi^L, \chi^R$}
    \RightLabel{\text{ Cut$^1$}}
    \BinaryInfC{$\Gamma, \chi^R$}
    \AxiomC{$\A_2''$}
    \UnaryInfC{$\chi^L, \psi^R$}
    \AxiomC{$\B$}
    \UnaryInfC{$\psi^L, \Delta$}
    \RightLabel{\text{ Cut$^2$}}
    \BinaryInfC{$\chi^L, \Delta$}
    \RightLabel{\text{ Cut$^3$}}
    \BinaryInfC{$\Gamma, \Delta$}
    \DisplayProof\\[2ex]
\end{tabular}
\end{center}
Either $\A_1''$ or $\A_2''$ is an axiom. In both cases, Cut$^3$ is of rank 2 and its cut formula is part of an axiom. The proofs ending with Cut$^1$ and Cut$^2$ are of strictly smaller size than the original proof, so by induction they can be made to satisfy the desired properties 1 and 2 (one of them is already rank 1).

The above cases cover all possibilities for how the premises of the topmost cut in the proof are constructed, concluding the proof.
\end{proof}

%\subsection{Subformula Property}

Since sequent calculus for orthologic has no elimination rule, traversing a sequent $S$ backwards in its proof obtained by \autoref{thm:cutelim} we obtain the following subformula property.
\begin{cor}[Subformula Property for Orthologic]
\label{cor:subformula}
If a sequent $S$ has a proof in the proof system of \autoref{fig:proofSystem} with axioms, then it has such a proof where each formula in each sequent ocurring in the proof is a subformula of  $S$ or a subformula of an axiom.
\end{cor}
\begin{proof}
Let $S$ be a sequent that has a proof. By \autoref{thm:cutelim}, consider a proof of $S$ with properties (1) and (2). We view the proof as a directed tree with root $S$, whose nodes are the sequents occurring in the proof. We show that each formula in the node is a subformula of $S$ or of an axiom, by induction on the distance of the tree node to the root $S$. The property trivially holds for the root $S$. Suppose now that the property holds for a node $S'$ in the tree; we show that it also holds for its children. We consider all applicable rules in~\autoref{fig:proofSystem}. Hyp rule has no children, so there is nothing to show. Consider a Cut rule and one of its children, $\Gamma, \psi^R$. Then $\Gamma$ is a subformula because it is a part of $S'$, whereas $\psi$ is a part of the axiom by property (1). The case for the other child $\psi, \Delta$ of the Cut rule is analogous. The case of the Left and Right rules are easy: we observe that their premises are sequents whose formulas are subformulas of the conclusion $S'$, so they are a subformula of $S'$ or of axioms, by inductive hypothesis and the transitivity of the ``subformula'' relation. Finally, if $S'$ is obtained using the axiom rule, then its formulas are subformulas of the axioms by definition.
\end{proof}

\section{Cubic-Time Proof Search}

\newcommand{\cmnt}[1]{\tcp*[r]{\footnotesize #1}}
\newcommand{\cmnts}[1]{\tcp*[f]{\footnotesize #1}}
\begin{algorithm}[hbt]
\DontPrintSemicolon
\caption{Cubic-Time Proof Search for $\OL$ with Axioms}\label{alg:pseudocode}
    \textbf{type} Formula$^*$ \cmnt{an annotated formula or a special None value}
    A: Set[(Formula$^*$, Formula$^*$)] $\gets$ Input \cmnt{set of axioms} 
    AxFormulas: Set[Formula] $\gets \bigcup \{ \{ a, b \} \mid \{ a^\Box, b^\circ \} \in \mbox{A} \}$
                                     \cmnt{formulas from axiom sequents} 
    proven: Set[(Formula$^*$, Formula$^*$)] $\gets$ Set.empty \cmnt{formulas already proven}
    visited: Set[(Formula$^*$, Formula$^*$)] $\gets$ Set.empty \cmnt{formulas whose proof is being attempted}  
    \procedure{prove($\Gamma$: Formula$^*$, $\Delta$: Formula$^*$)}{
        \lineIf{proven.contains(($\Gamma$, $\Delta$))}{\Return \True}
        \lineElseIf{visited.contains(($\Gamma$, $\Delta$))}{\Return \False}
        \Else{
        visited.add(($\Gamma$, $\Delta$))\cmnt{necessary to avoid infinite loops}
        success $\gets$ $\lbrace$\;\Indp
            ($\Gamma$==None \&\& prove($\Delta$, $\Delta$) ) || ($\Delta$==None \&\& prove($\Gamma$, $\Gamma$) ) || \cmnt{Impicit contraction}
            (($\Gamma$==$\phi^L$ \&\& $\Delta$ == $\phi^R$) || ($\Gamma$==$\phi^R$ \&\& $\Delta$ == $\phi^L$))  || \cmnt{Hyp}
            (($\Gamma, \Delta)\in$ A)  || \cmnt{Ax($\Gamma$, $\Delta$)}
            ($\Gamma$!=None \&\& $\Delta$!=None \&\& (prove($\Gamma$, None) || prove(None, $\Delta$))) || \cmnt{Weaken}
            ($\Gamma$==$(\neg\phi)^L$ \&\& prove($\phi^R$, $\Delta$)) || \cmnt{LeftNot}
            ($\Gamma$==$(\phi\land\psi)^L$ \&\& (prove($\phi^L$, $\Delta$) || prove($\psi^L$, $\Delta$)) || \cmnt{LeftAnd}
            ($\Gamma$==$(\phi\lor\psi)^L$ \&\& prove($\phi^L$, $\Delta$) \&\& prove($\psi^L$, $\Delta$) || \cmnt{LeftOr}
            ... || \cmnt{analogous Right cases for $\Gamma$}
            ... || \cmnt{analogous Left and Right cases for $\Delta$}
            AxFormulas.exists((x:Formula) $\rightarrow$ \;
                \ \ \ \  prove($\Gamma$, x$^R$) \&\& prove(x$^L$, $\Delta$) || prove($\Delta$, x$^R$) \&\& prove(x$^L$, $\Gamma$)) \cmnt{Cut with axiom}
        \Indm$\rbrace$\;
        \If{success}{
            %visited.remove(($\Gamma$, $\Delta$))\;
            proven.add(($\Gamma$, $\Delta$))}
        \Return success      
        }
    }
\end{algorithm}

In this section we show that the subformula property (Corollary~\ref{cor:subformula}) not only implies a quadratic bound on the size of proofs, but also allows us to define an $O(n^3)$ proof search procedure. Such deterministic polynomial-time search is in contrast to co-NP completeness of validity in classical propositional logic \cite{10.1145/800157.805047} and to PSPACE completeness of validity in intuitionistic logic \cite{DBLP:journals/tcs/Statman79, DBLP:conf/tlca/Urzyczyn97}. We assume that the set $A$ of axioms is finite throughout this section.

Our approach is to eliminate from the proof system in \autoref{fig:proofSystem} deduction steps which do not satisfy the conditions of \autoref{thm:cutelim}. We assume that no axiom is a trivial one $a \le a$, as that one does not help prove anything. 

For a formula, sequent or set of formulas or sequents $o$, let $\sizeOf{o}$ denote the number of subformulas in $o$. This is asymptotically equal to the number of symbols needed to represent $o$. In particular for the set of axioms, $\sizeOf{A}$ is the number of subformulas in all axioms of $A$ whereas $|A|$ is the number of axioms.

\begin{lemma}[Bound on Intermediate Sequents]
\label{lem:subformulaCount}
There is a function that when given $S$ computes a set $\relevantS{S}$ containing at most $4(\sizeOf{S}+\sizeOf{A})^2$ intermediate sequents such that if $S$ is provable then it is provable with a proof whose all sequents appear in $\relevantS{S}$.
\end{lemma}
\begin{proof}    
    By Corollary~\ref{cor:subformula}, there is a proof where each formula is a subformula of $S$ or of $A$. Let $\relevantS{S}$ be the set of all sequents built from these formulas. There are at most $2(\sizeOf{S}+\sizeOf{A})$ of labelled subformulas. A sequent has at most two labelled subformulas, and the number of sequents is bounded by the number of ordered pairs of labelled formulas, which is $4(\sizeOf{S}+\sizeOf{A})^2$.
\end{proof}

\begin{lemma}[Bound on Branching]
\label{lem:proofcount}
    Let $A$ be a set of axioms, and let $S$ be a sequent. Then,  if $S$ has a proof in the system of~\autoref{fig:proofSystem} then there are at most $7+4|A|$ valid instances of rules whose conclusion is $S$. We call the sequents above $S$ in these steps the \textit{possible parents} of $S$.
\end{lemma}
\begin{proof}
Let $S$ be a sequent of the form $\Gamma, \Delta$. We state in parentheses the maximal number of valid instances for each of the rules in \autoref{fig:proofSystem}:
\begin{itemize}
    \item(1) $S$ can be deduced by an application of the hypothesis rule or from the axiom rule (not both as we assume the axioms are non-trivial)
    \item (2) $\Gamma, \Delta$ can follow using weakening, from either $\Gamma$ or $\Delta$
    \item (2) $\Gamma, \Delta$ can follow using a Left or a Right rule on $\Gamma$ in at most two ways, depending on the structure of $\Gamma$:
    \begin{itemize}
        \item If $\Gamma = (\neg \phi)^L$ then $S$ can be deduced from $\phi^R, \Delta$ with LeftNot
        \item If $\Gamma = (\phi \land \psi)^L$ then $S$ can be deduced in two ways using LeftAnd: from $\phi^R, \Delta$ and from $\psi^R, \Delta$
        \item If $\Gamma = (\phi \lor \psi)^L$ then $S$ can be deduced in exactly one way using LeftOr: from both  $\phi^R, \Delta$ and $\psi^R, \Delta$.
        \item Right rules are symmetrical and either Left or Right rules apply to $\Gamma$, not both
    \end{itemize}
    \item (2) $\Gamma, \Delta$ can follow using a Left or Right rule on $\Delta$ in also at most two ways, entirely analogously to the previous case.
    \item (4$|A|$) $\Gamma, \Delta$ can be deduced using the Cut rule in $4|A|$ different ways: the cut formula $\phi$ can be any formula among the left or right sides of axioms (thanks to \autoref{thm:cutelim} property (1)), for at most $2|A|$ different formulas, and for each the Cut instance can be either of
    \begin{center}
\begin{tabular}{c c c}
    \AxiomC{$\Gamma, \phi^R$}
    \AxiomC{$\phi^L, \Delta$}
    \RightLabel{\text{ Cut}}
    \BinaryInfC{$\Gamma, \Delta$}
    \DisplayProof &
    \hspace{1em} or \hspace{1em} &
    
    \AxiomC{$\Delta, \phi^R$}
    \AxiomC{$\phi^L, \Gamma$}
    \RightLabel{\text{ Cut}}
    \BinaryInfC{$\Gamma, \Delta$}
    \DisplayProof
    \\[2ex]
\end{tabular}
\end{center}    
\end{itemize}
We thus obtain the desired bound $1 + 2 + 2 + 2 + 4|A| = 7 + 4|A|$.
\end{proof}

\begin{thm}
\label{thm:proofsearch}
    There is a proof search procedure for $\OL$ running in time $\mathcal{O}((1+|A|)n^2)$, where $|A|$ is the number of given axioms and $n$ the total size of the problem.
\end{thm}
\begin{proof}
In \autoref{alg:pseudocode} we present pseudocode for backward search. Each line from 8 to 16 corresponds to trying a specific deduction rule.

For an input sequent $S$, the proof strategy consists in recursively computing the possible parents of $S$. By \autoref{lem:subformulaCount}, proof of a sequent $S$ need only involve at most $4(\sizeOf{S}+\sizeOf{A})^2$ intermediate sequents, which is $\mathcal{O}(n^2)$.
Moreover, note that to compute the possible parents of a sequent, we need  not observe the formulas entirely but only their roots, so computing one possible parent is constant time. Moreover, by \autoref{lem:proofcount}, a sequent can only have at most $C+4|A|$ possible parents, so reducing a sequent to all of its possible parents has complexity $\mathcal{O}(1+|A|)$. The final complexity of the proof search procedure is bounded by generating parents for all possible sequents we can encounter, which is $\mathcal{O}((C+|A|)n^2) =
\mathcal{O}((1+|A|)n^2)$. 
\end{proof}

\paragraph{Note on the complexity of memoization}
The complexity argument of the previous proof requires memoization to be efficient. However, a naive implementation of the ``visited'' and ``proven'' sets can increase the total runtime if the sets are implemented as lists or rely on structural equality of formulas, which has itself complexity linear in the size of the formulas. Instead, we can use the following approach. Assign to every annotated formula a unique integer (of size $\mathcal{O}(\log(n))$, or a word in the usual word RAM model). Then assign a field to every annotated formula, storing a hash table $m$ with integers as keys and Booleans as values. When a sequent of the form $a,b$ is proven, set a.m(b.id) and b.m(a.id) to True. If the proof search fails, set them to False. 
With this representation, the checks of lines 7-10 reduce to one access in a hash table, and line 24 to one update (which takes amortized constant time in the word RAM model).

\subsection{Merging Axioms for Quadratic Complexity}

In the particular case where $A=\emptyset$, \autoref{alg:pseudocode} is quadratic, which is also the best known result for the word problem and normalization problem in both ortholattices and lattices \cite{whitmanFreeLattices1941a, guilloudFormulaNormalizationsVerification2023}. 
In general, it can be beneficial to keep the number of axioms as small as possible. For this purpose, we can combine axioms with the same left-hand side into one. Given two axioms representing $a \le b_1$ and $a \le b_2$ we can merge them into an equivalent one $a \le b_1 \land b_2$. Indeed, given an axiom sequent $\{a^L, (b_1 \land b_2)^R\}$ we can derive $\{a^L, b_1^R\}$ as follows:
\begin{center}  
        \AxiomC{}
        \RightLabel{ Ax}
        \UnaryInfC{$a^L, (b_1 \land b_2)^R$}

        \AxiomC{}
        \RightLabel{ \hypname}
        \UnaryInfC{$b_1^L, b_1^R$}
        \RightLabel{LeftAnd}
        \UnaryInfC{$(b_1 \land b_2)^L, b_1^R$}
        \RightLabel{Cut}

        \BinaryInfC{$a^L, b_1^R$}

        \DisplayProof
\end{center}
Dually, we can merge axioms with the same right-hand side, $a_1 \le b$ and $a_2 \le b$ into $a_1 \lor a_2 \le b$. Finally, $a \le \lnot b$ can be rewritten into $b \le \lnot a$ and vice versa. We can repeat this process until all left-hand sides and all right-hand sides of axioms are distinct, and no left side is a complement of a right side (we can even use normal forms for ortholattices to make such checks more general~\cite{guilloudFormulaNormalizationsVerification2023}). Such axiom pre-processing transformations do not change the set of provable formulas. They can be done in time $\mathcal{O}(n^2)$ and they reduce $|A|$ while not increasing $n$. Using such transformations can thus improve the cubic bound for certain kinds of axiom sets. 
As a very special case, if all axioms have the form $1 \le b_i$ and $a_i \le 0$ (corresponding to singleton sequents), we can combine them into a single axiom, obtaining $\mathcal{O}(n^2)$ complexity.
\begin{cor}
    There is an $\mathcal{O}(n^2)$ algorithm for checking provability from axiom sets $A$ in which all axioms are singleton sequents.
\end{cor}

\section{Proof Strength of Orthologic with Axioms}
The previous section established a cubic time algorithm (\autoref{alg:pseudocode}) for deriving all consequences of axioms that hold in orthologic. This generalization is sound for classical logic while still being efficient. A key question then is: how precise is it as an approximation of classical logic? 

To help answer this question, we present several classes of classical problems that our algorithm solves \emph{exactly}: it is not only sound for them (as it is for all problems), but also \emph{complete}: it always finds a proof if, e.g., a SAT solver would find it. Furthermore, we partly characterize our $\OL$ proof system in terms of restricted forms of resolution for propositional logic.

We are interested in traditional classes of deduction problems that are solvable by $\OL$ proofs. Formally, we define the deduction problem in orthologic, and, respectively, classical logic.
\begin{defin}
\label{defin:deductionproblem}
    An instance of the deduction problem is characterized by a pair $(A, S)$ where $A$ is a set of axioms and $S$ the goal, all of which are sequents whose interpretation as inequality is given by \autoref{sec:proofsystem}. The deduction problem in $\OL$ (resp. $\CL$) consists in deciding if the goal $S$ can be derived from axioms $A$ in orthologic (resp. classical logic). If this is the case, then the instance is called \textit{valid}.
\end{defin}

\begin{defin}
    An instance of the deduction problem is $\OL$-solvable if and only if it has the same validity in $\OL$ and $\CL$. A class of instances of the deduction problem is $\OL$-solvable if and only if all its members are $\OL$-solvable.
\end{defin}
\noindent As $\OL$ is sound relative to $\CL$, 
the following are equivalent formulations of $\OL$-solvability:
\begin{itemize}
    \item The instance has a proof in $\OL$ if and only if it has a proof in $\CL$.
    \item The goal of the instance is true in all ortholattice interpretations satisfying the axioms if and only if it is true in all $\lbrace 0,1\rbrace$ interpretations satisfying the axioms. 
\end{itemize}
In particular, if the goal of the instance is the empty sequent (hence, we talk about the consistency of axioms), $\OL$-solvability of $(A,\emptyset)$ is equivalent to each of the following statements:
\begin{itemize}
    \item The axioms of the instance are either unsatisfiable in $\OL$ or satisfiable in $\CL$. 
    \item The axioms of the instance either have a non-trivial Boolean model, or admit only the trivial one-element structure as a model among all ortholattices.
\end{itemize}
In particular, \autoref{thm:proofsearch} gives a polynomial-time decision procedure with respect to classical logic for any class of deduction problems instances that are $\OL$-solvable.

In the sequel, we look at the satisfiability of propositional logic formulas in conjunctive normal form (CNF), which are conjunctions of clauses, as their analysis plays an important role in proof theory of $\CL$ and the practice of SAT solving.
%While the satisfiability problem on any formula in $\CL$ reduces efficiently to the satisfiability of a formula in CNF via the Tseitin transform \cite{tseitinComplexityDerivationPropositional1983}, this requires the distributivity law and does not work in $\OL$. %Hence, CNF problem are only a strict subset of all deduction problems we can consider. 
%We will see however in \autoref{sec:tseitin} that $\OL$ admits a flattening procedure similar (but not equivalent) to the Tseitin transform.
Among the simplest and most studied refutationally complete systems for $\CL$ is resolution on clauses, shown in \autoref{fig:resolution}. 
    \begin{figure}[ht]
\centering
        \begin{tabular}{l l}
            \multicolumn{2}{c}{
                \AxiomC{$C, x$}
            \AxiomC{$C', \neg x$}
            \RightLabel{\text { Resolution}}
            \BinaryInfC{$C, C'$}
            \DisplayProof 
            }\\[2ex]
            \AxiomC{}
            \RightLabel{\text { Hypothesis}}
            \UnaryInfC{$C, x, \neg x$}
            \DisplayProof&
            \AxiomC{$C$}
            \RightLabel{\text { Weaken}}
            \UnaryInfC{$C, C'$}
            \DisplayProof
        \end{tabular}    
\caption{The Resolution proof system with hypothesis and weakening rules. $C$ and $C'$ represent arbitrary sets of literals. This system is complete for deriving contradictions in $\CL$, with formulas expressed in conjunctive normal form, even without the Weaken and Hypothesis rules \cite[Chapter 2]{robinsonHandbookAutomatedReasoning2001}. }
\label{fig:resolution}
\end{figure}

\subsection{Completeness for 2SAT}     
\label{ex:2sat}

We start with the simplest example of a CNF, the 2CNF class.
    A 2CNF formula is a finite set of clauses $C_1,...,C_m$, where each clause is a disjunction of two literals or a single literal. 
    For example, $(x \lor \neg y), (\neg x \lor z), (\neg z)$ is a 2CNF formula. 2SAT if the problem of deciding if a 2CNF formula is satisfiable, i.e. if it has a model in the two-element Boolean algebra. Conversely, the instance is unsatisfiable if and only if the conjunction of the clauses implies falsity.
    
    We next show how to encode a 2SAT instance into an $\OL$ deduction problem. 
    The idea is to view a 2SAT instance as a deduction problem $(A, \emptyset)$ where
    each axiom in $A$ is a sequent containing at most two labelled \emph{variables} as formulas.
    We create an axiom sequent for each clause, where a negative literal $\lnot p$ becomes labelled formula $p^L$ and a positive literal $p$ becomes $p^R$. For example,  
    $\{ \neg x, y\rbrace$ becomes $x^L, y^R$. Similarly, 
    $\{x, y\}$ becomes $x^R, y^R$, whereas $\{ x \}$ becomes $x^R$.
    This encoding is equivalent (in $\CL$) to the 2SAT instance, with the interpretation of sequents given in \autoref{defin:interpretation}.

    Consider the Resolution prof system shown in  \autoref{fig:resolution}. For 2SAT instances, the outcome of resolution can be simulated by orthologic using the Cut rule, which allows us to prove the following.
\begin{thm}
    2SAT is $\OL$-solvable.
\end{thm}
\begin{proof}
    Consider a 2CNF and its representation as a set of sequents. The instance is unsatisfiable in $\CL$ if and only if there exist a derivation of the empty clause in Resolution. We proceed by induction on the Resolution derivation to show that if a clause is derived, then there is an $\OL$-proof of the corresponding sequent.

    Consider a Resolution step between two clauses of (at most) two elements. The sequent corresponding to its conclusion can be deduced from the sequent corresponding to its premises by a single application of the Cut rule in orthologic. 

    \begin{center}        
    \begin{tabular}{c c c}
        \AxiomC{$\lbrace \gamma, y \rbrace$}
        \AxiomC{$\lbrace \neg y, \delta \rbrace$}
        \RightLabel{\text{ Resolution}}
        \BinaryInfC{$\gamma, \delta$}
        \DisplayProof &
        
        $\hookrightarrow$ &
        
        \AxiomC{$\lbrace \Gamma, y^R \rbrace$}
        \AxiomC{$\lbrace y^L, \Delta \rbrace$}
        \RightLabel{\text{ Cut}}
        \BinaryInfC{$\Gamma, \Delta$}
        \DisplayProof
    \end{tabular}
    \end{center}
    Where $\gamma$ and $\delta$ are one arbitrary literals (or no literal) and $\Gamma$ and $\Delta$ represent the corresponding annotated formula (or absence thereof).
    Weaken and Hypothesis steps are similarly simulated by the eponymous steps in $\OL$.
    %(and in fact, using LeftAnd and RightOr, for arbitrary clauses and not only 2CNF). %Hence, the $\OL$ proof system simulates Resolution proofs in problems with clauses of cardinality at most 2.
    
    Conversely, if the empty sequent is derivable in orthologic, then no non-trivial ortholattice satisfy the assumptions, and hence no Boolean algebra.
\end{proof}

\subsection{Orthologic Emulates Unit Resolution}

For an arbitrary clause $N \cup P$, let it be encoded as the sequent $(\bigwedge N)^L, (\bigvee P)^R$, where $N$ (resp. $P$) is the set of negative (resp. positive) variables in the clause. A \emph{Unit Resolution} step is a Resolution step (\autoref{fig:resolution}) where $C=\emptyset$ or $C'=\emptyset$.
    Interpreted over sequents, the application of a Unit Resolution step on two clauses $C_1=\lbrace x_i \rbrace$ and $C_2=(\lbrace \neg x_1,\ldots,\neg x_n\rbrace \cup P)$ corresponds to the deduction rule UnitResolutionR. Dually, for $C_1 = \{ \lnot x_i \}$ we obtain UnitResolutionL:
    \begin{center}
        \AxiomC{$(x_i)^R$}
        \AxiomC{$(x_1\land \ldots \land x_{i-1} \land x_i \land x_{i+1} \land \ldots \land x_n)^L, (\bigvee P)^R$}
        \RightLabel{\text{ UnitResolutionR}}
        \BinaryInfC{$(x_1 \land \ldots \land x_{i-1} \land x_{i+1} \land\ldots\land x_n)^L, (\bigvee P)^R$}
        \DisplayProof 

\vspace*{2ex}

        \AxiomC{$(x_i)^L$}
        \AxiomC{$(\bigwedge N)^L, (x_1\lor\ldots\lor x_{i-1}\lor x_i\lor x_{i+1}\lor\ldots\lor x_n)^R$}
        \RightLabel{\text{ UnitResolutionL}}
        \BinaryInfC{$(\bigwedge N)^L, (x_1\lor\ldots\lor x_{i-1}\lor x_{i+1}\lor\ldots\lor x_n)^R$}

        \DisplayProof 
    \end{center}
\begin{lemma}[$\OL$ simulates Unit Resolution] \label{lem:unitResolution}
  UnitResolutionL and UnitResolutionR are admissible rules (\autoref{defin:admissible}) for $\OL$ proof
  system of \autoref{fig:proofSystem}.
\end{lemma}
\begin{proof}    
    %Note that while the meet and join are technically binary operation, thanks to the associativity law allows us to write $(\bigwedge N)$ and $(\bigvee P)$ without ambiguity.
    We aim to show that this step is admissible in $\OL$ proofs. Instead of giving a syntactic transformation, we can see that the steps are sound in $\OL$ with a short semantic argument. 
    For UnitResolutionL,
    consider any ortholattice and assume that the premises of the rules are true. 
    The meaning of $x_i^R$ is $1 \le x^R$, which implies $x_i = 1$. This implies 
    $(x_1\land ...\land x_i \land ...\land x_n) = (x_1\land \ldots \land 1 \land \ldots x_n)$, 
    so the value of the non-unit premise clause reduces to the truth of the conclusion of the rule.    
    By completeness (\autoref{lem:soundCompPropOl}), there exists a proof of the conclusion from the premises. Thus, UnitResolutionR is admissible. The argument for UnitResolutionL is dual.
    \end{proof}
%\noindent A syntactic example of such a Unit Resolution rule as an orthologic proof can be found in the proof of \autoref{lem:substterms}.

\subsection{Completeness for Horn Clauses}

A Horn clause is a disjunction of literals such that at most one literal is positive. We encode a Horn clause into a sequent as
$(a_1\land ... \land a_n)^L, b^R$
where $a_i$ are the negated variables of the clause (if any), and $b$ the positive literal of the clause (if it exists). %Clauses without any such $a_i$ are called \textit{facts}.
A Horn instance is a conjunction of Horn clauses, and is unsatisfiable if and only if the empty clause can be deduced from it using Resolution. 

We encode a Horn instance into a deduction problem by adding an axiom for each Horn clause, and the empty sequent as the goal.

Horn instances can be solved using Unit Resolution only \citep{minouxLTURSimplifiedLineartime1988}. By \autoref{lem:unitResolution}, $\OL$ is complete for Horn instances. Other classes solvable using only Unit Resolution include \textit{q-Horn}, \textit{extended Horn} and \textit{renamed Horn} instances \cite{cepekKnownNewClasses2005}.
    
\begin{cor}[\textbf{Horn Clause Completeness}]
    \label{ex:horn}
    Horn Clause instances, q-Horn instances, extended Horn instances, and renamed Horn instances are
    $\OL$-solvable classes.
\end{cor}
Note that, despite our use of semantic techniques to show completeness for resolution, the results of \autoref{thm:proofsearch} apply, provide polynomial-time guarantees for solving these instances.

Renamed Horn instances are an interesting extension of Horn instances. A conjunction of clauses is renamed Horn if and only if there exists a set of variables $V$ such that complementing variables of $V$ in the instance yields a Horn instance. 
In particular, a clause in a renamed Horn instance can contain multiple positive and negative literals.  Unit Resolution is stable under such renaming, meaning that a unit resolution derivation of the empty clause remains a valid unit resolution derivation of the empty clause after renaming. This is the reason that Unit Resolution is also complete for renamed Horn clauses.

\subsection{Renaming Deduction Problems}

Motivated by renamed Horn instances, we now consider renaming of general deduction problems. We show that renamings of $\OL$-solvable instances are $\OL$-solvable.

\begin{defin}
    For an arbitrary variable $x$, the complement of $x$ is $\neg x$ and the complement of $\neg x$ is $x$. Two deduction problems $I_1$ and $I_2$ are \textit{renamings} of each other if there exists a set of variables $V$ such that complementing variables of $V$ in the axioms and goal of $I_1$ yields $I_2$.
\end{defin}

\begin{lemma}
    Let $I$ be a deduction instance such that $I$ is valid in $\OL$ if and only if it is valid in $\CL$. Then all renamed versions of $I$ are valid in $\OL$ if and only if they are valid in $\CL$.
\end{lemma}
\begin{proof}
    In $\OL$, if $I_1$ and $I_2$ are renamed versions of each other by a set of variables $V$, $I_1$ and $I_2$ have the same validity. Indeed, suppose there is an ortholattice $\mathcal O$ and assignement $s_1:V\rightarrow \mathcal O$ that is a counter model of $I$ (meaning it satisfies the axioms but not the goal, so $I_1$ is invalid). Then define $s_2$ such that $s_2(x)=s_1(x)$ if $x\not\in V$ and $s_2(x)=\neg s_1(x)$ if $x \in V$. It is then easy to check that since $\neg \neg x = x$ in $\OL$, $s_2$ is a counter model for $I_2$. Hence if $I_1$ is invalid then $I_2$ is invalid. Conversely, if $I_2$ admits a counter model, then $I_1$ admits a counter model as well.
    
    Similarly in $\CL$, $I_1$ and $I_2$ have the same validity by the same argument, considering assignments $V \rightarrow \{0, 1\}$ as counter models. 
    Hence, since $I_1$ has the same validity in $\OL$ and $\CL$, so does $I_2$.

\end{proof}

\subsection{Tseitin's Transformation for Orthologic Axioms}

Tseitin's transformation for classical logic transforms a formula with arbitrary alternations of conjunctions and disjunctions into one in Conjunctive Normal Form (CNF) that is equisatisfiable. The transformation works by introducing a linear number of new variables, and runs in near linear time. This justifies the focus of SAT solvers on solving formulas in CNF. 

The essence of Tseitin's transformation in classical logic is to introduce variables, such as $x$, that serve as names for subformulas, such as $F$, with the interpretation of $x$ bound to be equal the interpretation of $F$. Unfortunately, we cannot express such transformation as a single $\OL$ formula, because knowing that $\lnot x \lor F$ equals $1$ inside a formula does not imply $x \le F$. On the other hand, once we make use of the power of axioms, the usual Tseitin's transformation again becomes possible. This shows the importance of adding axioms to orthologic reasoning.

\begin{defin}[$\boldsymbol{\OL}$-Tseitin transformation]
Given a deduction problem instance with axioms $A$ and goal $S$ (where we assume without loss of generality that all formulas are in negation normal form), pick an arbitrary strict subexpression $e$ of a formula in $A$ or $S$ of the form $x \land y$ or $x \lor y$, for some literals $x$ and $y$. Pick a fresh variable $c$, introduce the axioms $c^L, e^R$ and $e^L, c^R$ and replace $e$ by $c$ in $A$ and $S$. Repeat until all formulas have height at most 2. We say that a problem $(A,S)$ is in Tseitin normal form if it is obtained from another problem using this transformation.
    
\end{defin}
The transformation does not alter the validity of a deduction problem instance in $\OL$, since we merely defined aliases for subexpressions. Indeed, having both $(c^L, e^R)$ and $(e^L, c^R)$ as axioms is equivalent to forcing $e=c$ in all models.

\subsection{Resolution Width for Orthologic Proofs in CNF}

Consider an $\OL$ proof for a deduction problem to which we applied $\OL$-Tseitin's transformation.
If $\circ$ and $\Box$ denote arbitrary $\{\_^L, \_^R\}$ annotations,
the resulting problem will only contain sequents of the form:
$\{a^\circ, (b \land c)^\Box\}$,\ 
$\{a^\circ, (b \lor c)^\Box\}$, \
$\{a^\circ, b^\Box\}$, \
$\{a^\circ\}$,\ and $\emptyset$,
for some literals $a$, $b$ and $c$.
%Moreover, $x \vdash y \lor z$ is strictly equivalent to having both $x \vdash y $ and $x \vdash z$, and similarly for 
Moreover, remember that by the subformula property (\autoref{cor:subformula}), if the problem admits a proof, then it has a proof that only uses formulas among $a$, $a\land b$ and $a\lor b$ (for any literals $a$ and $b$ appearing in the problem).
Hence, we can constrain every proof of $S$ to involve only sequents of at most 4 literals. In classical logic, every sequent appearing in the proof would then be equivalent to a conjunction of disjunctions of literals (i.e. a conjunction of clauses).  In the simplest case:
$$(w\land x)^L,  (y \lor z)^R \hspace{1em} \leadsto \hspace{1em} (\neg w \lor \neg x \lor y \lor z)$$
For sequents involving a left disjunction or right conjunction, using greatest lower bound and lowest upper bound properties of $\land$ and $\lor$:
\begin{align*}
(w\lor x)^L, (y \lor z)^R \hspace{1em} &\leadsto \hspace{1em} (\neg w \lor y \lor z) \land (\neg x \lor y \lor z)\\
(w\land x)^L, ( y \land z)^R \hspace{1em} &\leadsto \hspace{1em} (\neg w \lor \neg x \lor y) \land (\neg w \lor \neg x \lor z)\\
(w\lor x)^L, (y \land z)^R \hspace{1em} &\leadsto \hspace{1em} (\neg w \lor y) \land (\neg w \lor z) \land (\neg x \lor y) \land (\neg x \lor z)
\end{align*}
And similarly with all combinations of $^L$ and $^R$, where, for example, a conjunction with $L$ polarity behaves much like a disjunction with $R$ polarity.
We say that such a set of clauses \textit{represents} the corresponding sequent. Crucially, each of these clauses contains at most 4 literals.

We now consider again the Resolution proof system of \autoref{fig:resolution}.
Note that we work with plain resolution and \emph{not} extended resolution that introduces fresh variables on the fly \cite{tseitinComplexityDerivationPropositional1983}.
\begin{defin}
    The \textbf{width} of a resolution proof is the number of literals in the largest clause appearing in the proof.
\end{defin}
The next theorem characterizes the width of those classical logic resolution proofs that suffice to establish all formulas provable by $\OL$ derivations. 
\begin{thm}
    An $\OL$ proof of a problem in $\OL$-Tseitin normal form can be simulated by \autoref{fig:resolution} proofs of width $5$. 
    %Moreover, there exists $\OL$ proofs whose conclusion cannot be deduced using such proofs of width $4$.
\end{thm}
\begin{proof}
Consider an $\OL$ proof that satisfies properties of \autoref{thm:cutelim}.
We proceed by induction on the structure of such proof. The cases of non-cut rules are immediate. Namely, Hypothesis and Weakening in $\OL$ are directly simulated by the corresponding steps in Resolution.
In LeftNot, the principal formula must be a literal, so that the clause representation of the conclusion is the same as the interpretation of the premise.
In LeftOr, similarly, the representation of the conclusion is the conjunction of the representations of the premises. LeftAnd is simulated in a Resolution proof by Weakening. Right- steps are symmetrical to Left-steps.
The only non-trivial case is the Cut rule. The cut formula can have different shape
    \begin{enumerate}
        \item The cut formula is a literal
        \begin{center}
            \AxiomC{$\Gamma, x^R$}
            \AxiomC{$x^L, \Delta$}
            \RightLabel{\text{ Cut}}
            \BinaryInfC{$\Gamma, \Delta$}
            \DisplayProof
        \end{center}
        We then have technically 36 different cases to describe depending on whether $\Gamma$ and $\Delta$ are conjunctions, disjunctions or literals and their polarity. We show the two extreme cases, and all other can be deduced by symmetry.

        If $\Gamma$ and $\Delta$ are left conjunctions, right disjunctions or literals, the Cut rule is simulated with a single Resolution instance:
\begin{center}
\begin{tabular}{c c c}
    \AxiomC{$(a\land b)^L, x^R$}
    \AxiomC{$x^L, (c\lor d)^R$}
    \RightLabel{\text{ Cut}}
    \BinaryInfC{$\Gamma, \Delta$}
    \DisplayProof&
    
    $\hookrightarrow$ &
    
    \AxiomC{$\lbrace \neg a, \neg b, x \rbrace$}
    \AxiomC{$\lbrace \neg x, c, d\rbrace$}
    \RightLabel{\text{ Resolution}}
    \BinaryInfC{$\lbrace\neg a, \neg b, c, d \rbrace$}
    \DisplayProof
\end{tabular}
\end{center}

If $\Gamma$ and $\Delta$ are left disjunctions or right conjunctions, 2 applications of Resolution are necessary to obtain each of the two clauses in the conclusion (4 if both):

\begin{center}
\begin{tabular}{c}
    \AxiomC{$(a\lor b)^L, x^R$}
    \AxiomC{$x^L, (c\land d)^R$}
    \RightLabel{\text{ Cut}}
    \BinaryInfC{$\Gamma, \Delta$}
    \DisplayProof\\
    
    $\hookrightarrow$ \\
    
    \AxiomC{$\lbrace \neg a, x\rbrace, \lbrace \neg b, x\rbrace$}
    \AxiomC{$\lbrace  \neg x, c \rbrace, \lbrace  \neg x, d\rbrace$}
    \RightLabel{\text{ $4 \times$ Resolution}}
    \BinaryInfC{$\lbrace\neg a, c\rbrace, \lbrace\neg a, d\rbrace, \lbrace\neg b, c\rbrace, \lbrace\neg b, d\rbrace$}
    \DisplayProof    
\end{tabular}
\end{center}
Where each of the clause in the conclusion can be reached by using Resolution on two of the clauses in the premises.
\item Consider now the case where the Cut formula is a conjunction (the disjunction case is symmetrical). 
    \begin{center}
            \AxiomC{$\Gamma, (x \land y)^R$}
            \AxiomC{$(x \land y)^L, \Delta$}
            \RightLabel{\text{ Cut}}
            \BinaryInfC{$\Gamma, \Delta$}
            \DisplayProof
        \end{center}
    We again present the two extreme cases. If $\Gamma$ and $\Delta$ are both left conjunctions or right disjunctions:
\begin{center}
\begin{tabular}{c}
    \AxiomC{$(a\land b)^L, (x \land y)^R$}
    \AxiomC{$(x \land y)^L, (c\lor d)^R$}
    \RightLabel{\text{ Cut}}
    \BinaryInfC{$(a\land b)^L, (c\lor d)^R$}
    \DisplayProof\\[3ex]
    
    $\hookrightarrow$\\[1ex]
    
    \AxiomC{$\lbrace \neg a, \neg b, x \rbrace$, $\lbrace \neg a, \neg b, y \rbrace$}
    \AxiomC{$\lbrace c, d, \neg x, \neg y\rbrace$}
    \RightLabel{\text{ Resolution on $x$}}
    \BinaryInfC{$\lbrace\neg a, \neg b, c, d, \neg y\rbrace$}
    \RightLabel{\text{ Resolution on $y$}}
    \UnaryInfC{$\lbrace\neg a, \neg b, c, d \rbrace$}
    \DisplayProof\\[4ex]
\end{tabular}
\end{center}
The conclusion can be reached by applying Resolution twice successively, but here the intermediate clause $\lbrace\neg a, \neg b, c, d, \neg y\rbrace$ reaches width $5$.
%Moreover, it is clear that the only other way to reach the conclusion from the axioms is to first resolve on $x$ and then on $y$. So, the conclusion cannot be reached by Resolution of width less than 5, and the bound is tight.

If $\Gamma$ and $\Delta$ are left disjunctions or right conjunctions:
\begin{center}
\begin{tabular}{c}
    \AxiomC{$(a\lor b)^L, (x \land y)^R$}
    \AxiomC{$(x \land y)^L, (c\land d)^R$}
    \RightLabel{\text{ Cut}}
    \BinaryInfC{$(a\lor b)^L, (c\land d)^R$}
    \DisplayProof\\[3ex]
    
    $\hookrightarrow$\\[1ex]
    
    \AxiomC{$\lbrace \neg a, x \rbrace$, $\lbrace \neg a, y \rbrace, \lbrace \neg b, x \rbrace$, $\lbrace \neg b, y \rbrace$}
    \AxiomC{$\lbrace \neg x, \neg y, c\rbrace, \lbrace \neg x, \neg y, d\rbrace$}
    \RightLabel{\text{ $4\times$ Resolution on $x$}}
    \BinaryInfC{$\lbrace \neg y, \neg a, c \rbrace, \lbrace \neg y,  \neg a, d \rbrace, \lbrace \neg y,  \neg b, c \rbrace, \lbrace \neg y,  \neg b, d \rbrace$}
    \RightLabel{\text{ $4\times$ Resolution on $y$}}
    \UnaryInfC{$\lbrace \neg a, c \rbrace, \lbrace \neg a, d \rbrace, \lbrace \neg b, c \rbrace, \lbrace \neg b, d \rbrace$}
    \DisplayProof\\[4ex]
\end{tabular}
\end{center}
the Cut can be simulated by first resolving 4 times on $x$ and then 4 times on $y$.
\end{enumerate}
Hence, Resolution of width 5 can simulate all $\OL$ proofs.
\end{proof}

%\textbf{TODO:} Resolution where all clauses have bounded width is polynomial time because the number of clauses to consider is polynomial, see also \cite{DBLP:conf/focs/Galil74}.
% https://dl.acm.org/doi/pdf/10.1145/800116.803755

\section{Effectively Propositional Orthologic}
\label{sec:epr}
So far we have studied propositional orthologic. In this section we introduce decidable classes of \emph{predicate} orthologic.
Our inspiration is the Bernays–Schönfinkel-Ramsey (BSR) class of classical first-order logic formulas \cite[Section 6.2.2]{BoergerETAL97ClassicalDecisionProblem}, which consists of formulas of first order logic that contain predicates and term variables but no function symbols, and whose prenex normal form is of the form
$
\exists x_1,...x_n. \forall y_1,...,y_n. \phi
$
where $\phi$ is quantifier free. Syntactically, a formula in the BSR class can be represented as a quantifier-free formula with constants and variables symbols, where the variables are implicitly universally quantified. It is also called \textit{Effectively Propositional Logic}  (EPR) \cite{piskacDecidingEffectivelyPropositional2010}, because deciding the validity of such formula can be reduced to deciding the validity of a formula in propositional logic by a grounding process. This is possible because formulas in the BSR class have finite Herbrand universe \cite[p.1798]{robinsonHandbookAutomatedReasoning2001}. BSR class (with its multi-sorted logic generalization) has found applications in verification \cite{DBLP:journals/pacmpl/PadonLSS17}.

We show that orthologic also admits a similar grounding process, allowing us to define the class of \emph{effectively propositional orthologic}. The class is EXPTIME in the worst case, in contrast to co-NEXPTIME for BSR. Moreover, we show that it becomes polynomial if we restrict the maximal number of variables in axioms, which is in contrast to the corresponding restriction yielding an NP-hard class for classical logic \cite[Section 6.2.2]{BoergerETAL97ClassicalDecisionProblem}.
%As in the previous sections, our formulas have the syntax of classical logic (in this case, its EPR fragment), but we will examine their semantics in ortholattices, subsequently identifying cases where the two semantics coincide.

In this section, \textit{variables} denote variable symbols at the term level, and not propositional variables. 
We fix two disjoint countably infinite sets of symbols: constants $C$, and variables $V$. A predicate signature $\Sigma$ specifies a finite set of predicate symbols $\{p_1,\ldots, p_n\}$ with their non-negative arities $s_i$, and a finite non-empty set of constants $\Bar{C} = \{c_1,\ldots,c_n\}$. Define the set of atomic formulas over $\Sigma$ as
\[
    P_\Sigma = \bigcup_{i=1}^n \{ p_i(\vec x) \mid \vec x \in (V \cup \Bar{C})^{s_i} \}
\]
An EPR formula (over $\Sigma$) is a formula constructed from $P_\Sigma$ using $\land,\lor,\lnot$, corresponding to $\term{\OL}{P_\Sigma}$. An annotated EPR formula is $a^L$ or $a^R$ where $a$ is an EPR formula. An EPR sequent is a set of at most two annotated EPR formulas. 
The degree of a formula or sequent is the number of distinct free variables in it. The degree of a finite set $A$ of sequents, $d(A)$, is the maximum of degrees of its sequents.
An atomic formula, formula, or a sequent is \emph{ground} if it contains no variables (only constants), that is, it has degree zero.
\begin{defin}
An EPR deduction instance is a set $(A,S)$ where $A$ (the axioms) is a set of EPR sequents and $S$ (the goal) is an EPR sequent.
\end{defin}
\begin{defin}
An instance of a formula (respectively, sequent) is a formula (or sequent) obtained by replacing all occurrences of some variables by other variables or constants. 
The expansion of a formula $s$ (respectively, sequent), denoted $s^*$, is the set of all of its instances. $s^*$ is countable, and  infinite if $s$ contains at least one variable. If $A$ is a set of sequents then $A^* = \bigcup \{ s^* \mid s \in A \}$.
For an EPR deduction instance $(A,S)$ its expansion is $(A^*,S)$.
\end{defin}
\begin{defin}
$\EPROLD$ is the problem, given a signature $\Sigma$ and EPR deduction instance $(A,S)$ over $\Sigma$, decide whether its expansion $(A^*,S)$ is a valid $\OL$ deduction instance (in the sense of \autoref{defin:deductionproblem}).
\end{defin}

%We call \textbf{Uniform Effectively OL Solving} the following problem.  
%    Let be given $p_1,...,p_n$ a set of predicate symbols each of arity $s_i$ and define $P = \lbrace p_i(\vec{x}) | i \leq n, \Vec{x} \in (V\cup C)^{s_i} \rbrace$ to be atoms. Let also be given a set of axioms $A = \lbrace a_1,...,a_m\rbrace$ with each $a_i$ being an inequality between two elements of $\term{P}{\OL}$. Define an \textit{instance} of $a_i$ to be $a_i$ with any number of variables substituted by another variable or constant. We note $A^*$ the set of all instances of all axioms in $A$. 
    
%Finally, let's be given a sequent $S = (\Gamma, \Delta)$ over $\term{P}{\OL}$. Decide if $S$ holds in all ortholattices generated by $P$ where all inequalities of $A^*$ hold.
    
%This amounts to asking if $S$ follows from axioms of the form $\exists^*\forall^*...$, where existentially quantified symbols correspond to constants and universally quantified ones correspond to free variables in our definitions.

    We first show as an intermediate lemma that we only need to look at instances of $A$ using variables appearing in $S$ and constants in $S$ and $A$.
\begin{lemma} \label{lem:varsNeeded}
    Suppose $S$  has a proof $\mathcal P$ involving axioms in $A^*$. Then it has a proof containing only variables that appear in $S$ and constants in $S$ and $A$.
\end{lemma}
\begin{proof}
    If a variable $z$ appears somewhere in $\mathcal{P}$ but not in $S$, then it has to be eliminated by a Cut rule at some point.
\begin{center}
\vspace{-1em}
\begin{tabular}{c c c}
    \AxiomC{$\A$}
    \UnaryInfC{$\Gamma, \phi(z)^R$}
    \AxiomC{$\B$}
    \UnaryInfC{$\phi(z)^L, \Delta$}
    \RightLabel{\text{ Cut}}
    \BinaryInfC{$\Gamma, \Delta$}
    \DisplayProof
\end{tabular}
\end{center}
Let $c \in \Bar{C}$ be any constant symbol of $\Sigma$. Let $\A[z:=c]$ be the proof $\A$ with every instance of $z$ replaced by $c$. For any given axiom $a \in A^*$, $a[z:=c]$ is also an axiom of $A^*$, so that all axioms steps occurring in $\A[z:=c]$ are correct. It is easy to see that all non-axioms steps in $\A[z:=c]$ remain correct under the substitution. Because by assumption the Cut rules eliminates $z$ from the formula, $\Gamma$ does not contain $z$ and hence the conclusion of $\A[z:=c]$ is exactly $\Gamma, \phi(c)^R$.
The same can be done to $\B$, so that we obtain a proof where $z$ does not appear:
\begin{center}
\begin{tabular}{c c c}
    \AxiomC{$\A[z:=c]$}
    \UnaryInfC{$\Gamma, \phi(c)^R$}
    \AxiomC{$\B[z:=c]$}
    \UnaryInfC{$\phi(c)^L, \Delta$}
    \RightLabel{\text{ Cut}}
    \BinaryInfC{$\Gamma, \Delta$}
    \DisplayProof 
\end{tabular}
\vspace{-0.8em}
\end{center}
To eliminate constant, we use the same argument except that since axioms are not stable under renaming of constants (but all other rules and in particular the hypothesis rule are), we cannot eliminate constant symbols appearing in $A$.
\end{proof}

\begin{thm}
\label{thm:EPOsolving}
    The $\EPROLD$ problem $(A,S)$ of size $n$ and degree $d(A)$ is solvable in $\textit{PTIME}(n^{d(A)})$.
\end{thm}

\begin{proof}
    We will reduce $\EPROLD$ to propositional $\OL$ deduction problems, which \autoref{alg:pseudocode} can then solve. By completeness of $\OL$ (\autoref{lem:soundCompPropOl}), whether $S$ holds in all ortholattices where $A$ holds is equivalent to the following question:
    Does there exist a finite subset $A'$ of $A^*$ such that $S$ has an orthologic proof with axioms among $A'$?

\autoref{lem:varsNeeded} implies that for any sequent $S$, we only need to consider a finite number of axioms, namely the axioms involving variables in $S$ and constants in $A$ and $S$ (minimum one). Each axiom $a$ has at most $(|S|+||A||)^{d(A)}$ such instances, so the total number of axiom we need to consider is $\mathcal{O}(|A|\cdot(|S|+||A||)^{d(A)}) = \mathcal{O}(n^{d(A)+1})$. Combining this result with \autoref{thm:proofsearch} gives $\textit{PTIME}(n^{d(A)})$, or, more precisely, $\mathcal{O}(n^{3(d(A)+1)})$.
\end{proof}

\subsection{Instantiation as a Rule}
Instead of starting by grounding all axioms, we can delay instantiation until later in the proof, yielding shorter proofs in some cases. Formally, we add an instantiation step to the proof calculus of \autoref{fig:proofSystem} over $\term{P}{\OL}$:

\begin{center}
\vspace{-0.5em}
\begin{tabular}{c c c}
    \AxiomC{$\Gamma, \Delta$}
    \RightLabel{\text{ Inst}}
    \UnaryInfC{$\Gamma[\vec{x}:=\vec{t}], \Delta[\vec{x}:=\vec{t}]$}
    \DisplayProof
\end{tabular}
\end{center}
Holding for arbitrary sets of term variable $\vec{x}$ and terms $\Vec{t}$. We note the resulting proof system $\OL^{I}$.

%\begin{defin}
%    
%\end{defin}
\begin{lemma}
    \label{lem:OlIadmissible}
    For a sequent $S$ and set of axioms $A$ over quantifier-free predicate logic, $S$ has has an $\OL$ proof from $A^*$ if and only if $S$ has an $\OL^{I}$ proof from $A$.
\end{lemma}

\begin{proof}\hspace{1em}\\
\vspace{-1.5em}
    \begin{itemize}
        \item[$\rightarrow$]: Given an $\OL$ proof with axioms in $A^*$, said axioms can be obtained from $A$ in $\OL^I$ by an application of the instantiation rule:
\begin{center}        
    %\vspace{2em}
\begin{tabular}{c c c}
    \AxiomC{}
    \RightLabel{\text{ Ax}}
    \UnaryInfC{$\Gamma^*, \Delta^*$}
    \UnaryInfC{$...$\vspace{2em}}
    \UnaryInfC{$S$}
    \DisplayProof &

    $\hookrightarrow$ &
    \AxiomC{}
    \RightLabel{\text{ Ax}}
    \UnaryInfC{$\Gamma, \Delta$}
    \RightLabel{\text{ Inst}}
    \UnaryInfC{$\Gamma^*, \Delta^*$}
    \UnaryInfC{$...$\vspace{5em}}
    \UnaryInfC{$S$}
    \DisplayProof
\end{tabular}
\end{center}
        Where $(\Gamma, \Delta) \in A$ and $(\Gamma^*, \Delta^*) \in A^*$.

        \item[$\leftarrow$]: Note that if the Inst rule are only uses right after axioms, then we can reverse the transformation above. We show that given a proof in $\OL^I$ using Inst, the instances of Inst can be swapped with other rules and be pushed to axioms. For example

\begin{center}
\begin{tabular}{c c c}
    \AxiomC{$\A'$}
    \UnaryInfC{$\alpha^L, \Delta$}
    \AxiomC{$\A''$}
    \UnaryInfC{$\beta^L, \Delta$}
    \RightLabel{\text { LeftOr}}
    \BinaryInfC{$(\alpha \lor \beta)^L, \Delta$}
    \RightLabel{\text { Inst}}
    \UnaryInfC{$(\alpha^* \lor \beta^*)^L, \Delta^*$}
    \DisplayProof &
    
    \hspace{1em} $\hookrightarrow$ \hspace{1em} &
    
    \AxiomC{$\A'$}
    \UnaryInfC{$\alpha^L, \Delta$}
    \RightLabel{\text { Inst}}
    \UnaryInfC{$(\alpha^*)^L, \Delta^*$}
    \AxiomC{$\A''$}
    \UnaryInfC{$\beta^L, \Delta$}
    \RightLabel{\text { Inst}}
    \UnaryInfC{$(\beta^*)^L, \Delta^*$}
    \RightLabel{\text { LeftOr}}
    \BinaryInfC{$(\alpha^* \lor \beta^*)^L, \Delta^*$}
    \DisplayProof
    \\[5ex]
\end{tabular}
\end{center}

    The cases for all other rules are similar. Then, any conclusion of an Inst rule is a member of $\A^*$ and can be replaced by an Axiom rule to obtain an $\OL$ proof from $A^*$.

    \end{itemize}
\vspace{-1.4em}
\end{proof}

\subsection{Proof Search with Unification}
While searching for a proof, we usually want to delay decision-making (such as which variable to instantiate) for as long as possible. In backward proof search, this means we want to delay it until the sequent is an axiom of $A^*$. In forward proof search, however, being able to use the Inst rule allows delaying instantiation as much as possible, as in resolution for first-order logic classical \cite{DBLP:journals/jacm/Robinson65}, \cite[Chapter 2]{robinsonHandbookAutomatedReasoning2001}. 

We thus adopt unification to decide when and how to instantiate a variable, whenever we use a rule with two premises. The corresponding directed rules are shown in \autoref{fig:UnifRules}. Without function symbols, the most general unifier of $\phi$ and $\psi$ is the substitution $\theta$ of smallest support such that $\theta(\phi) = \theta(\psi)$ \cite[Chapter 8]{robinsonHandbookAutomatedReasoning2001}.
\iffalse
\begin{figure}[ht]
    \begin{center}
        \begin{tabular}{l}
            \AxiomC{$C, a$}
            \AxiomC{$C', \neg b$}
            \RightLabel{\text { Resolution}}
            \BinaryInfC{$\theta(C), \theta(C')$}
            \DisplayProof 
        \end{tabular}
    \end{center}
    \caption{The resolution rule for first order logic, with unification. $\theta$ is the most general unifier of $a$ and $b$}
    \label{fig:resolutionFOL}
\end{figure}
\fi

\begin{figure}[ht]
    \begin{center}
        \begin{tabular}{l l}
            \multicolumn{2}{c}{
                \AxiomC{$\Gamma, \phi^R$}
                \RightLabel{\text{ Inst}}
                \UnaryInfC{$\theta(\Gamma), \theta(\phi)^R$}
                \AxiomC{$\psi^L, \Delta$}
                \RightLabel{\text{ Inst}}
                \UnaryInfC{$\theta(\psi)^L, \theta(\Delta)$}
                \RightLabel{\text{ Cut}}
                \BinaryInfC{$\theta(\Gamma), \theta(\Delta)$}
                \DisplayProof
            }\\[5ex]
            \multicolumn{2}{c}{
                where $\theta$ is the most general unifier of $\phi$ and $\psi$
            }\\[2ex]

            \AxiomC{$\Gamma_1, \phi^R$}
                \RightLabel{\text{ Inst}}
                \UnaryInfC{$\theta(\Gamma_1), \theta(\phi)^R$}
            \AxiomC{$\Gamma_2, \psi^R$}
                \RightLabel{\text{ Inst}}
                \UnaryInfC{$\theta(\Gamma_2), \theta(\psi)^R$}
            \RightLabel{\text{ RightAnd}}
            \BinaryInfC{$\theta(\Gamma_1), \theta(\phi \land \psi)^R$}
            \DisplayProof &
        
            \AxiomC{$\Gamma_1, \phi^L$}
                \RightLabel{\text{ Inst}}
                \UnaryInfC{$\theta(\Gamma_1), \theta(\phi)^R$}
            \AxiomC{$\Gamma_2, \psi^L$}
                \RightLabel{\text{ Inst}}
                \UnaryInfC{$\theta(\Gamma_2), \theta(\psi)^R$}
            \RightLabel{\text{ LeftOr}}
            \BinaryInfC{$\theta(\Gamma_1), \theta(\phi \lor \psi)^L$}
            \DisplayProof
            \\[5ex]
            \multicolumn{2}{c}{
                where $\theta$ is the most general unifier of $\Gamma_1$ and $\Gamma_2$
            }

        \end{tabular}
    \end{center}
\caption{Sequent-calculus style deduction rules with unification for Effectively Propositional Orthologic.}
\label{fig:UnifRules}
\end{figure}

\begin{thm}
    A sequent $S$ over $\term{P}{\OL}$ has a proof in $\OL^I$ if and only if there exists a sequent $S'$ such that $S$ is a particular instantiation of $S'$ and  $S'$ has a proof where the Inst rule is only used in the specific cases of \autoref{fig:UnifRules} or to rename variables.
\end{thm}

\begin{proof} (Sketch). The proof is once again by induction and case analysis on the proof $\mathcal P$ of $S$, except we move the instantiation step toward the conclusion of the proof.
Any instance of the Inst rule in $\mathcal P$ can be swaped with the next rule, unless it is a Cut, LeftOr or RightAnd rule. Consider the proof rule that follows the instantiation:
\begin{itemize}
    \item Hypothesis is a leaf rule, so it can never follow a step.
    \item Weaken is immediate, as long as the variables in $\Delta$ are properly renamed
\begin{center}
\begin{tabular}{c c c}
    \AxiomC{$\A$}
    \UnaryInfC{$\Gamma$}
    \RightLabel{\text { Inst}}
    \UnaryInfC{$\sigma(\Gamma)$}
    \RightLabel{\text { Weaken}}
    \UnaryInfC{$\sigma(\Gamma), \Delta$}
    \DisplayProof &
    
    \hspace{1em} $\hookrightarrow$ \hspace{1em} &
    
    \AxiomC{$\A$}
    \UnaryInfC{$\Gamma$}
    \RightLabel{\text { Weaken}}
    \UnaryInfC{$\sigma(\Gamma), \pi(\Delta)$}
    \RightLabel{\text { Inst}}
    \UnaryInfC{$\sigma(\Gamma), \Delta$}
    \DisplayProof 
    \\[5ex]
\end{tabular}
\end{center}
        Where $\pi$ is a renaming of variables in $\Delta$ to names that are fresh. In particular, it is invertible.
    \item LeftNot and RightNot are immediate.
    \item For RightOr and LeftAnd, the transformation is the same as Weaken.
    \item In the Cut case, assume that the two premises $(\Gamma, \phi^R)$ and $(\psi^L, \Delta)$ have no shared variables, by renaming them if necessary:
\begin{center}
\begin{tabular}{c c c}
    \AxiomC{$\Gamma, \phi^R$}
    \RightLabel{\text{ Inst}}
    \UnaryInfC{$\sigma_1(\Gamma), \sigma_1(\phi)^R$}
    \AxiomC{$\psi^L, \Delta$}
    \RightLabel{\text{ Inst}}
    \UnaryInfC{$\sigma_2(\psi)^L, \sigma_2(\Delta)$}
    \RightLabel{\text{ Cut}}
    \BinaryInfC{$\sigma_1(\Gamma), \sigma_2(\Delta)$}
    \DisplayProof \\[5ex]
    
    $\hookrightarrow$ \\
    
    \AxiomC{$\Gamma, \phi^R$}
    \RightLabel{\text{ Inst}}
    \UnaryInfC{$\theta(\Gamma), \theta(\phi)^R$}
    \AxiomC{$\psi^L, \Delta$}
    \RightLabel{\text{ Inst}}
    \UnaryInfC{$\theta(\psi)^L, \theta(\Delta)$}
    \RightLabel{\text{ Cut}}
    \BinaryInfC{$\theta(\Gamma), \theta(\Delta)$}
    \RightLabel{\text{ Inst}}
    \UnaryInfC{$\sigma_1(\psi)^L, \sigma_2(\Delta)$}
    \DisplayProof
    \\[6ex]
\end{tabular}
\end{center}
Note that since $\theta$ is the most general unifier for $\phi$ and $\psi$, and $\sigma_1(\phi) = \sigma_2(\psi)$, $\theta$ factors in both $\sigma_1$ and $\sigma_2$, so that the last Inst step is correct.
    \item LeftOr and RightAnd are similar to the Cut step.
\end{itemize}
\vspace{-1em}
\end{proof}
%In practice, in a forward proof search, formally renaming variables for each sequent is not needed. We can simply always consider variables in different sequents to be unique.

\subsection{Solving and Extending Datalog Programs with Orthologic}

Datalog is a logical and declarative programming language admitting formulas in a further restriction of the BSR class where $\phi$ is forced to be a Horn clause (over predicates). A Datalog program is then a conjunction of such formulas (or clauses) \cite{ullmanPrinciplesDatabaseKnowledgeBase1988, ullmanPrinciplesDatabaseKnowledgeBase1989, dantsinComplexityExpressivePower2001}. While the validity problem for the BSR class is coNEXPTIME complete \cite{ramseyProblemFormalLogic1930, piskacDecidingEffectivelyPropositional2010}, solving a Datalog program is only EXPTIME-complete. This makes Datalog a suitable language for logic programming and database queries.
Typically, a Datalog query asks if a certain fact (an atom without variable) is a consequence of the clauses in the program. This naturally corresponds to solving a deduction problem, with the axioms corresponding to the program and the goal to the query.
\begin{lemma}
    Datalog program can be evaluated using orthologic.
\end{lemma}
\begin{proof}
    Datalog programs and queries form a subset of the BSR class. \autoref{thm:EPOsolving} shows that such problems can be reduced via grounding to purely propositional $\OL$. Moreover, as the resulting set of axioms contains only Horn clauses, \autoref{ex:horn} implies that the Datalog program has the same semantic in $\OL$ and in $\CL$.
\end{proof}
This means that for any Datalog program, the $\OL$ semantic agrees with the classical semantic. \autoref{alg:pseudocode} hence provides a decision procedure for Datalog with complexity $\textit{PTIME}(n^{d(A)})$ (as \autoref{thm:EPOsolving} shows).
This matches known complexity classes of Datalog, which has exponential query complexity (corresponding here to axioms) and polynomial data complexity (corresponding to the goal and axioms with no variables,  or \textit{facts}) \cite{dantsinComplexityExpressivePower2001}.

%and compare orthologic-based programming to other extensions of Datalog with negation and disjunctions.

\subsection{Axiomatizing Congruence and Equality Relations}

We next show that, when equality is axiomatized in effectively propositional orthologic, a substitution rule becomes admissible.
Let $X$ be a countably infinite set of term variables whose elements are noted $x,y,z,...$. Let be given a presentation with predicate symbols $p_1,..., p_n$ each of arity $s_i$ and $\sim$ a predicate symbol representing a congruence relation on $X$.
Consider the set of axiom $A_\sim$ containing the following sequents that axiomatize the equivalence property:
\begin{align*}
                               &(x\sim x)^R \\
                   (x\sim y)^L,&(y\sim x)^R \\
     (x\sim y \land y\sim z)^L,&(x\sim z)^R 
\end{align*}
and for each symbol $p_i$ and each $1 \leq j\leq s_i$, the congruence property for $p_i$:
$$
(x\sim y \land p_i(z_1,...,z_{j-1}, x, z_{j+1},...,z_{s_i}))^L, p_i(z_1,...,z_{j-1}, y, z_{j+1},...,z_{s_i})^R
$$

Again, this does not constitute a finite presentation of an ortholattice, as there are infinitely many possible instances of axioms.
However, by \autoref{thm:EPOsolving}, if a sequent $S$ over $X_\sim$ has a proof involving axioms in $A_\sim$, then it has a proof with only variables that appear in $S$. Moreover, the degree of $A_\sim$ is $d(A_\sim) = \max(3, \max_i(s_i)+1)$ axioms, so that the complexity of the proof search is exponential in the arity of the predicates in the language and polynomial in the size of the problem, for a fixed language.
The following lemma shows that in any decision problem whose axioms contain $A$, we can add a substitution rule for equality.
\begin{lemma}
Fix a set of predicate symbols and constants. Let $A$ be a set of axiom such that $\A_\sim \subset A$. The following rule for substitution of equal terms is admissible in $\OL$ with axioms in $A$:
\begin{center}
    \AxiomC{$\Gamma[x:=s], \Delta[x:=s]$}
    \AxiomC{$s\sim t$}
    \RightLabel{\textup { Subst$_\sim$}}
    \BinaryInfC{$\Gamma[x:=t], \Delta[x:=t]$}
    \DisplayProof 
\end{center}
\label{lem:substterms}    
\end{lemma}
\begin{proof}
    Suppose $x$ occurs only once in $\Gamma, \Delta$ (if it appears multiple time, we repeat the argument). Suppose without loss of generality that this unique occurrence is in $\Gamma$. Let $a(x) \equiv p_i(u_1,...,x,...u_{s_i})$ be the atomic formula containing this occurrence of $x$, i.e. $\Gamma = \Gamma'[\chi:=a(x)]$, for a propositional variable $\chi$. Axioms in $A_\sim$ allow the following proof, where we first derive $a(s)^L, a(t)^R$:
    \begin{center}
    %\resizebox{.9\linewidth}{!}{

        \AxiomC{$s\sim t^R$}
        \RightLabel{\text { Weak.}}
        \UnaryInfC{$a(s)^L, s\sim t^R$}
        \RightLabel{\text { RightAnd}}
        \AxiomC{}
        \RightLabel{\text { Hyp.}}
        \UnaryInfC{$a(s)^L, a(s)^R$}
        \RightLabel{\text { R.And}}
        \BinaryInfC{$a(s)^L, (s\sim t \land a(s))^R$}

        \AxiomC{}
        \RightLabel{\text { Ax.}}
        \UnaryInfC{$(s\sim t \land a(s))^L, a(t)^R$}
        \RightLabel{\text { Cut}}
        \BinaryInfC{$a(s)^L, a(t)^R$}
        
        \DisplayProof     
    \end{center}    
    and then conclude, using an analogous derivation of $a(t)^L, a(s)^R$
    \begin{center}
        \AxiomC{$\Gamma[x:=s], \Delta$}
        \dashedLine
        \UnaryInfC{$\Gamma'[\chi:= a(s)], \Delta$}

        \AxiomC{$a(s)^L, a(t)^R$}

        \AxiomC{$s\sim t^R$}
        \AxiomC{}
        \RightLabel{ Ax.}
        \UnaryInfC{$s\sim t^L, t\sim s^R$}
        \RightLabel{\text{ Cut}}
        \BinaryInfC{$t\sim s^R$}
        \UnaryInfC{\ldots(analogous)\ldots}
        \UnaryInfC{$a(t)^L, a(s)^R$}
        \RightLabel{\text{ Subst}}
        \TrinaryInfC{$\Gamma'[\chi:= a(t)], \Delta$}
        
        \dashedLine
        \UnaryInfC{$\Gamma[x := t], \Delta$}
        \DisplayProof     
    \end{center}
The dashed lines denote syntactic equality. Note that we have shown the Subst rule for propositions to be admissible in orthologic in~\autoref{lem:substformulas}. Hence, Subst$_\sim$ is admissible in orthologic with any axiomatization containing $A_\sim$.
\end{proof}

\iffalse
\textbf{TODO:} Please summarize as a theorem: what complexity do we get for labelledDatalog with congruence relations? In fact, in the absence of function symbols, we can just say that this suffices to axiomatize equality? Can we also use it to justify proof rule that substitutes equivalent things with equivalent things? The models would suggest yes, but in the absence of completeness we don't know. If not here, the substitution rule should be shown to hold for propositional case in the previous section, where we are certain that it holds because the proof system is sound and complete.
\fi %Done

%These examples illustrate that exact classical logic semantic has long been forgone in logic programming, in favor of more computational and procedural semantics. In fact, one could wonder what it means for a statement to be a consequence of a (logical) program if you have no practical way to deduce it. Orthologic is a natural candidate for a semantic  that is both logicaly pure (although not classical) and efficiently computable, and it is a strict extension of Datalog semantic with disjunction and negation. 

%As we have shown, equality can be axiomatized in $\OL$. \citep{zhangBetterTogetherUnifying2023} efforts to include efficient equality reasoning in Datalog have been using e-graphs and equality saturation, giving a more specialized and effective approach than \textbf{TODO}

%We hence conjecture that an orthologic-based logic programming framework with equality has potential.

%\section{Predicate Orthologic}
%\input{fool}

\section{Further Related Work}
% cite William McCunne: Automatic Proofs and Counterexample for Some Ortholattice Identities (as evidence that people did not have decision procedures but used ad-hoc techniques)

\citet{brunsFreeOrtholattices1976} first solved the word problem for free ortholattices in 1976 using algebraic techniques extending the work of \citet{whitmanFreeLattices1941a}, who first solved the word problem for free lattices.

The observation that we can obtain a proof system for Orthologic by restricting Gentzen's sequent calculus to sequents with at most two formulas was already made by \citet{schultemontingCutEliminationWord1981}. They also showed that the system without axioms admits Cut Elimination. \citet{eglyDifferentProofSearchStrategies2003} used the same system to describe a backward proof-search procedure exponential both in time and proof size, and a polynomial ($\Omega(n^7)$) forward procedure. We improved this result to a $\mathcal{O}(n^2)$ (time and size) backward proof search procedure.
Other proof systems have been considered. \citet{meinanderSolutionUniformWord2010} used a different set of inference rule to show that the word problem for finitely presented ortholattices is decidable in polynomial time. Their solution involves exhaustive forward deduction, and they give no precise exponent of the polynomial.

\citet{laurentFocusingOrthologic2016} introduces two other sequent-based proof systems for orthologic using concepts from linear logic, and in particular \textit{focusing}, to constrain proof search. Their procedure is forward-driven. While no complexity analysis is provided, their algorithm is clearly polynomial, and experimental benchmark shows improvement over the algorithm of \citep{eglyDifferentProofSearchStrategies2003}.

\citet{kawanoLabeledSequentCalculus2018} describes a proof system of a different flavour for orthologic without axioms. It is based on label sequents and designed to allow for an implication symbol. However, no complexity of proof search is given and the system does not have a limit on the number of formulas per sequent. This prevented the author from deriving a bound on the total number of sequents, as we do in \autoref{lem:proofcount}.

\citet{guilloudFormulaNormalizationsVerification2023} use ortholattices in the context of software verification as an approximation of Boolean algebra. They present an algorithm able to normalize any formula into an equivalent one (by ortholattices laws) of smallest size, with the goal of improving caching efficiency and reducing formula size for SMT solving. Their approach does not involve a proof system and does not support axioms in general.
%they use $\leq_\OL$ in practice to encode some relationship on atoms such as integers or arrays. Benchmarks on software verification show that being able to add non-logical axiom yields significant improvements.

Researchers have explored extensions of Datalog with negation and disjunction, with various computable semantics. Most of these models do not correspond to classical models of predicate logic. Datalog is declarative, but most of its extensions rely on procedural semantics.
In a line of work starting with \cite{clarkNegationFailure1978}, negation has been introduced with a failure meaning, where if $a$ cannot be verified, then $\neg a$ is taken to hold. 
%This is also the PROLOG approach. 
\cite{kunenNegationLogicProgramming1987} introduces a formal (but not decidable) semantic for negation as failure.
\cite{aptTheoryDeclarativeKnowledge1988} introduces the notion of stratified programs. This consists in specifying layers of clauses evaluated increasingly, so that if $a$ is not shown in a level, $\neg a$ is assumed in the next, which differs from orthologic semantics.
\cite{gelfondClassicalNegationLogic1991} introduce a notion of classical negation but the logic is not classical, as $\neg a \rightarrow b$ and $\neg b \rightarrow a$ are not equivalent in the proposed semantics.
The use of Datalog in program analysis inspired researchers to define Datalog with lattice semantics \cite{DBLP:conf/pldi/MadsenYL16}, which explicitly incorporates the concept of fixed point of particular lattices into the language semantics. Our approach is instead to view Datalog in the broader context of validity in all ortholattices, with orthologic as a convervative approximation of validity for classical logic that is always sound and, in several cases we identified, complete.

\section{Conclusions}
We have studied algorithmic and proof-theoretic properties of orthologic, a sound generalization of classical logic based on ortholattices.  We have shown a form of generalized cut elimination for propositional orthologic in the presence of axiom, implying a subformula property. We have used this result to design a cubic-time proof search procedure for orthologic with axioms (quadratic with bounded cardinality of axiom sets).
Furthermore, we have shown that some classes of classical decision problems including 2CNF, propositional Horn clause generalizations and Datalog always admit $\OL$ proofs. This provides sound and complete polynomial-time reasoning for a number of theorem proving tasks in classical logic. We anticipate applications of orthologic with axioms in predictable proof automation inside proof checkers, program verifiers, and expressive type systems.
%The ability to support lattice complement and disjunction inside a predictable Datalog algorithm opens up the possibility to build property-driven static program analyses that use both forward inference on reachable states and backward inference on the desired properties to establish.  
\begin{acks}
We thank Sankalp Gambhir for his helpful comments on a draft of this paper.
We thank anonymous reviewers of POPL 2024 for their helpful feedback.
\end{acks}

\raggedright
\bibliographystyle{ACM-Reference-Format}
\bibliography{sguilloud.bib,more.bib}

%\newpage
%\section{TO REMOVE}
%\input{remove.tex}

\end{document}